\documentclass[12pt, preprint]{aastex}
\newcommand\nspec[4]{\mbox{$#1\,^{#2}{#3}_{#4}$}}
\newcommand\zp{$\zeta$~Pup}
\newcommand\zo{$\zeta$~Ori}
\newcommand\io{$\iota$~Ori}
\newcommand\deltao{$\delta$~Ori}

\begin{document}

\title{Measurements and analysis of helium-like triplet ratios in the X-ray
  spectra of O-type stars}

\author{Maurice A.\,Leutenegger and Frits B. S.\,Paerels}
\affil{Department of Physics and Columbia Astrophysics Laboratory}
\affil{Columbia University, 550 West 120th Street, New York, NY 10027, USA}
\email{maurice@astro.columbia.edu}
\author{Steven M.\,Kahn}
\affil{Kavli Institute for Particle Astrophysics and Cosmology, Stanford
  Linear Accelerator Center and Stanford University}
\affil{2575 Sand Hill Road, Menlo Park, CA 94025}
\and
\author{David H.\,Cohen}
\affil{Department of Physics and Astronomy}
\affil{Swarthmore College, 500 College Avenue, Swarthmore, PA 19081}

\shorttitle{He-like triplet ratios in O stars}
\shortauthors{Leutenegger et al.}

\begin{abstract}

We discuss new methods of measuring and interpreting the
forbidden-to-intercombination line ratios of helium-like triplets in the X-ray
spectra of O-type stars, including accounting for the spatial distribution of
the X-ray emitting plasma and using the detailed photospheric UV
spectrum. Measurements are made for four O stars using archival {\it Chandra}
HETGS data. We assume an X-ray emitting plasma spatially distributed in the
wind above some minimum radius \(R_0\). We find minimum radii of formation
typically in the range of \(1.25 < R_0 / R_* < 1.67\), which is consistent
with results obtained independently from line profile fits. We find no
evidence for anomalously low \(f/i\) ratios and we do not require the
existence of X-ray emitting plasmas at radii that are too small to generate
sufficiently strong shocks. 

\end{abstract}


\section{Introduction}

Since the discovery of X-ray emission from OB stars by {\it Einstein}
\citep{Hel79, Sel79}, the exact mechanism for X-ray production has been
something of a mystery. X-ray emission from OB stars had been predicted by
\citet{CO79}, who proposed that an X-ray emitting corona could explain the
observation of superionized \ion{O}{6} through Auger ionization of
\ion{O}{4}. However, subsequent observations showing less attenuation of
soft X-rays than would be expected from a corona lying below a dense stellar
wind made a purely coronal origin seem unlikely \citep{CS83}. \citet{Mel93}
also found that a distributed X-ray source was necessary to explain the
observed \ion{O}{6} UV P Cygni profile in \zp. Furthermore,
with no expectation of a solar-type \(\alpha-\Omega\) dynamo in OB stars with
radiative envelopes, the coronal model fell out of favor. Subsequently,
several scenarios in which magnetic field generation and dynamos could exist
in OB stars have been proposed \citep{CM01, MC03, MM05}. Since these models
have been proposed, the primary observational evidence invoked by their
proponents is anomalously low $f/i$ ratios in the X-ray emission of a few
He-like ions in several stars. Re-examining these line ratios and determining
whether they require a coronal model to explain them is one of the main goals
of this paper. 

Shocks arising from instabilities in the star's radiatively driven
wind have been considered to provide a more likely origin for the observed
X-ray emission, as they are expected to be present, given the line-driven
nature of these winds \citep*{LW80, L82, KR85, OCR88, F95}. However, there have
been difficulties in reproducing the observed X-ray properties of O~stars,
such as the overall X-ray luminosity and the spectral energy distribution,
from stellar wind instability models \citep{Hel93, F95, FKPPP97, FPP97}. Until
recently, the quality of the available spectral data provided little insight
into these problems, since the CCD and proportional counter spectra could not
resolve individual spectral lines. 

Recent high resolution X-ray spectroscopy of OB stars by the XMM-{\it Newton}
Reflection Grating Spectrometer (RGS) \citep{Kel01, Mel03, RCMMT05} and the
{\it Chandra} High Energy Transmission Grating Spectrometer (HETGS)
\citep{SCHL00, WC01, CMWMC01, MCWMC02, Cel03, KCO03, Gel05, Cel06} have
answered some questions while raising new ones. Some stars have X-ray spectra
that appear consistent with emission from shocks in the wind, but the detailed
comparisons to predicted spectral models are still problematic. Both
\citet{WC01} and \citet{CMWMC01} have found low forbidden-to-intercombination
line ratios in one set of helium-like triplets each in the X-ray spectra of
\zo\, and \zp. They infer from this that some of the X-ray emitting plasma is
too close to the star to allow shocks of sufficient velocity to develop.

Other stars ($\theta^1$~Ori~C and $\tau$~Sco) have X-ray spectra that are
unusually hard and have relatively small line widths. While these stars might
be considered prime candidates for a coronal model of X-ray emission -
especially after having magnetic fields detected via Zeeman splitting
\citep{Del02, Del06} -  their behavior is better understood in terms of the
magnetically channeled wind shock model, rather than a model of magnetic
heating \citep{SCHL00, Cel03, SCHT03, Gel05, Del06}. 

Finally, we note that for all of the O giants and supergiants observed, the
line profiles are less asymmetric than predicted, given the high mass-loss
rates measured for these stars using radio free-free emission, H~$\alpha$
emission, and UV absorption lines \citep{WC01, Kel01, CMWMC01, MCWMC02, KCO03,
  Cel06}. This implies either a lower effective opacity to X-rays in their
winds (e.g. due to clumping or porosity effects \citep{FOH03, OFH04, OFH06,
  OC06}), or lower mass-loss rates \citep{Cel02, MFSH03, Hel03, BLH05, FMP06}.

One of the key diagnostic measurements available to us in understanding the
nature of X-ray emission in OB stars is the forbidden-to-intercombination line
ratio in the emission from ions that are isoelectronic with helium. This
ratio is sensitive to the UV flux, and thus to the proximity to the stellar
surface. This allows us to constrain the location of the X-ray emitting plasma
independently of other spectral data, such as emission line profile shapes.

In this paper we discuss methods for using the $f/i$ ratio to
constrain the location of X-ray emitting plasma in O star winds. In
particular, we explore the effects of a spatially distributed source motivated
by the broad line profiles. We discuss the effects of photospheric absorption
lines, as well as the $f/i$ ratio expected for a plasma emitted over a range
of radii, taking account of detailed line shapes when signal-to-noise
allows. We find that accounting in detail for photospheric absorption lines is
not important, as long as the X-ray emission originates over a range of
radii.

These methods are then applied to He-like triplet emission in a set of
archival {\it Chandra} observations of O stars. Our primary result is that
good fits can be acheived for most lines with models having emission
distributed over the wind, with minimum radii of about 1.5 stellar radii. We
find that none of the data require the X-ray emitting plasma to be formed very
close to the photosphere.

This paper is organized as follows: In \S~\ref{sec:model} we review the physics
of line formation in He-like species (\S~\ref{sub:fi}), explore the effects of
spectral structure in the photoexciting UV field (\S~\ref{sub:photo}), and of
spatial distribution of the X-ray emitting plasma (\S~\ref{sub:Rbar}), while
incorporating the line-ratio modeling into a self-consistent line-profile
model (\S~\ref{sub:hewind}). In \S~\ref{sec:data} we discuss the reduction and
analysis of archival O star X-ray spectra. In \S~\ref{sec:results} we give the
results of this analysis, fitting high signal-to-noise complexes with the
self-consistent line-profile model described in \S~\ref{sub:hewind} and
fitting the lower signal-to-noise complexes with multiple Gaussians and
interpreting these results according to the spatially distributed picture
described in \S~\ref{sub:Rbar}. In \S~\ref{sec:discussion} we discuss the
implications of these results, and in \S~\ref{sec:conclusions} we give our
conclusions.

\section{Model \label{sec:model}}

\subsection{Radial dependence of the $f/i$ ratio \label{sub:fi}}

The physics of helium-like ions in coronal plasmas has been investigated in
numerous papers \citep{GJ69, BDT72, GJ73, MS75, MS78a, MS78b, MS78c, PS81,
  P82, PMDRK01}. The principal diagnostic is the ratio of the strengths of the
forbidden to intercombination lines, \({\cal R} \equiv f/i\). We will use the
calligraphic \({\cal R}\) to refer to this ratio, and the italic $R$ to refer
to distances comparable to the stellar radius.

The upper level of the forbidden line (\nspec{2}{3}{S}{1}) is metastable and
relatively long-lived. When the excitation rate from \nspec{2}{3}{S}{1} to the
upper levels of the intercombination line (\nspec{2}{3}{P}{1,2}) becomes
comparable to the decay rate of the forbidden transition, the line ratio is
altered.\footnote{The \nspec{2}{3}{S}{1} state may also be excited to the
  \nspec{2}{3}{P}{0} state, but this state does not decay to ground, so we
  omit it from our discussion. However, in \citet{GJ69} and \citet*{BDT72},
  the formal treatment involves all states.} The excitations may be due to
electron impacts in a high density plasma, or due to an external UV radiation
source.

\citet{GJ69} (hereafter GJ) and \citet*{BDT72} (hereafter BDT) derive the
expression  
\begin{equation}
{\cal R} = {\cal R}_0\, \frac{1}{1 + \phi / \phi_c +
  n_e / n_c}
\end{equation}
where $\phi$ is the photoexcitation rate from \nspec{2}{3}{S}{} to
\nspec{2}{3}{P}{}, and \(\phi_c\) is the critical rate at which
${\cal R}$ is reduced to \({\cal R}_0/2\). Similarly, \(n_e\) is the
electron density, and \(n_c\) is the critical density. 

In Table~\ref{TabHeParams} we give our adopted values for the atomic parameters
necessary for calculation of He-like triplet ratios. We adopt the BDT values
for \(\phi_c\) because they have calculated it for all the ions we are
interested in, and because more recent calculations are not substantially
different. However, we use the more recent values for \({\cal R}_0\) from
\citet{PMDRK01}; their calculations of \({\cal R}_0\) are slightly lower than
those of BDT. \citet{PMDRK01} also give values for \(G \equiv (f + i) / r\),
evaluated at \(T_{max}\), the temperature at which emission from that He-like
ion is the strongest. We cite \(G(T_{max})\) for comparison with our
measurements.

Because densities high enough to cause a change in the line ratios exist only
very close to the star, we consider only the photoexcitation term.
If there are O stars with $f/i$ ratios that are measured to be too low to be
explained by photoexcitation, it is appropriate to consider the effects of
high density; this is not the case for any of our measurements.

The expression for $\phi$ may be evaluated as follows, given a model stellar
atmosphere Eddington flux \(H_{\nu}\): 
\begin{equation}
\phi = \frac{16 \pi^2 e^2}{m_e c}\, f\, \frac{H_{\nu}}{h \nu}\, W(r)
\end{equation}
where \(W(r) = \frac{1}{2}(1 - \sqrt{1 - (R_* / r)^2})\) is the geometrical
dilution.

The expression for the \({\cal R}\) ratio derived by GJ is written such that
$f$ is the sum of the oscillator strengths for \nspec{2}{3}{S}{1} to all three
of \nspec{2}{3}{P}{J}, despite the fact that \nspec{2}{3}{P}{0} does not decay
to ground, and \nspec{2}{3}{P}{2} only contributes for high $Z$. For low $Z$
(\ion{Ne}{9} and lower) \(H_{\nu}\) should be evaluated for
\nspec{2}{3}{S}{1} \(\rightarrow\) \nspec{2}{3}{P}{1}. For \ion{Mg}{11} and
higher $Z$ ions it is more accurate to evaluate \(H_{\nu}\) for both
\nspec{2}{3}{S}{1} \(\rightarrow\) \nspec{2}{3}{P}{1} and \nspec{2}{3}{P}{2}
and weight the average by the relative contributions to the effective
branching ratio. Of course, this is only necessary if \(H_{\nu}\) is
substantially different for the two transitions.

Since the flux of UV radiation seen by ions in a stellar wind decreases in
proportion to the geometrical dilution factor \(W(r)\), the \(\cal{R}\)
ratio is also a function of radius. It is helpful to express it in
this form:

\begin{equation}
{\cal R}(r) = {\cal R}_0 \,\frac{1}{1 + 2\, P\, W(r)}
\label{eqn:rofr}
\end{equation}

with \(P = \phi_* / \phi_c\) and

\begin{equation}
\phi_* = 8\pi\, \frac{\pi e^2}{m_e c}\, f\, \frac{H_{\nu}}{h\nu}.
\label{eqn:phistar}
\end{equation}
The value of the \({\cal R}\) ratio near the photosphere is then
\({\cal R}_{ph} = {\cal R}_0 / (1 + P)\).

In this paper we perform calculations and make measurements for a sample of
four O stars observed by {\it Chandra}: \zp, \zo, \io, and \deltao. The
relevant properties of these stars are given in Table~\ref{TabStarParams}. The
effective temperatures and gravities of the stars are taken from \citet{LL93}
and then rounded off to the closest values calculated on the TLUSTY O star
grid \citep{LH03}.

\subsection{The effect of photospheric absorption lines \label{sub:photo}}

The expression for \({\cal R}(r)\) written in the last paragraph involves an
approximation that must be explored further. We assumed a photospheric UV
flux that would be diluted by geometry, but we neglected the Doppler shift of
the absorbing ions. Over the range of Doppler shifts seen in a stellar wind,
there can be many photospheric absorption lines. This introduces an additional
radial dependence to the photoexcitation rate, and thus the \({\cal R}\) ratio:
\begin{equation}
\phi(r) \propto H_{\nu(r)} W(r)
\end{equation}
with the Doppler shifted frequency as seen by an ion at radius $r$:
\begin{equation}
\nu(r) = \nu_0 \left(1 + \frac{v(r)}{c}\right)
\end{equation}
In this expression a positive velocity respresents a blue shift.

\begin{figure}[p]
  \begin{center}
    \plotone{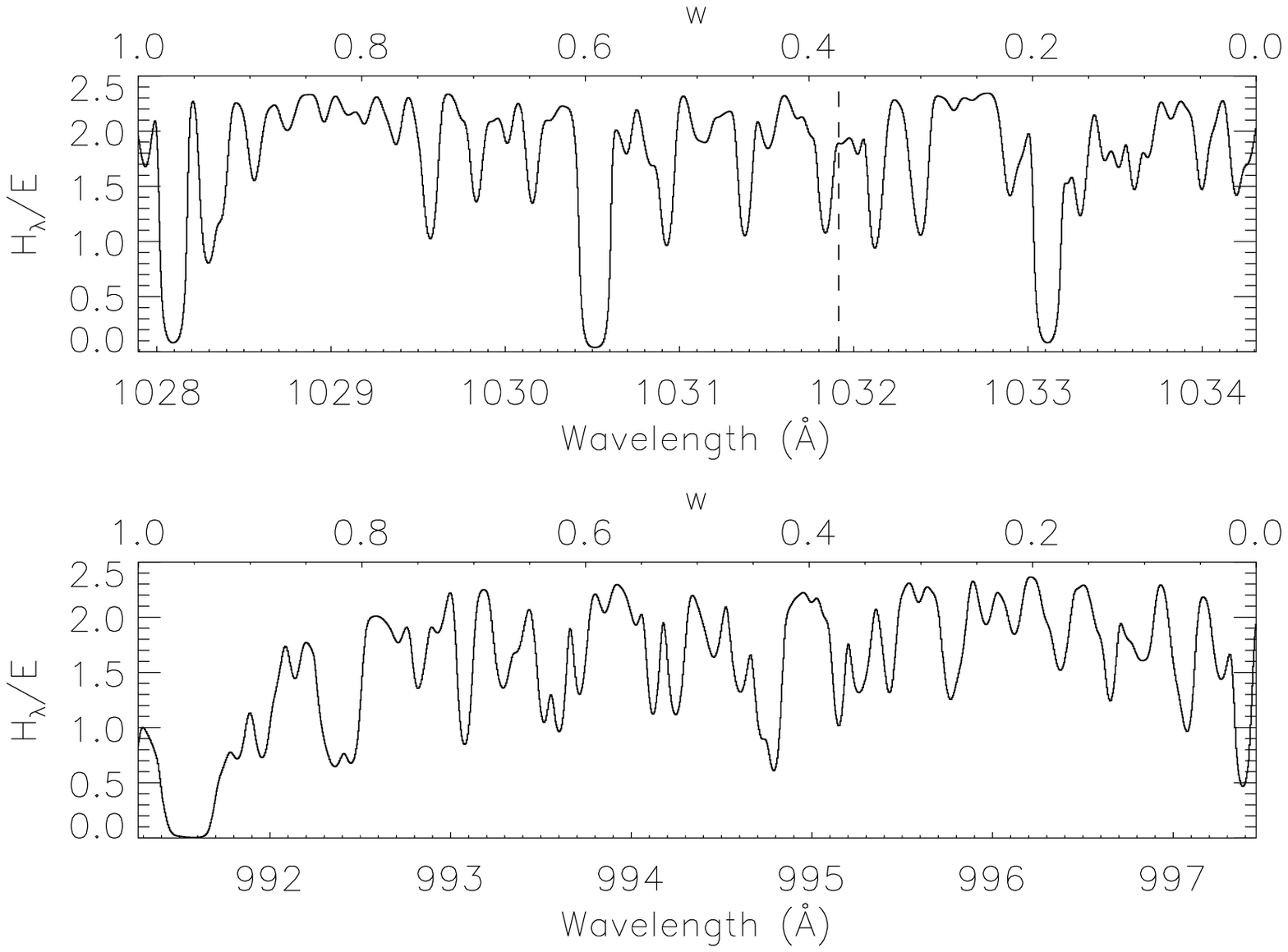}
  \end{center}
\caption{Model UV flux for \zo\, near the \nspec{2}{3}{S}{1} \(\rightarrow\)
  \nspec{2}{3}{P}{1,2} transitions of \ion{Mg}{11} (\(J=1\) is on the top,
  \(J=2\) on the bottom), plotted as a function of wavelength (bottom axis)
  and scaled stellar wind velocity, \(w(u)=v(u)/v_{\infty}\) (top axis). The
  flux is given in units of \(\mathrm{10^{20}\, photons\, cm^{-2}\, s^{-1}\,
    \AA^{-1}}\). The dashed line shows the rest wavelength of the \ion{O}{6}
  line at 1031.91 \AA. For comparison, the average continuum flux
  we use for this ion and this star is 1.67, in the same units. The model flux
  is taken from the TLUSTY O star grid \citep{LH03}.}
\label{tlusty}
\end{figure}

In Figure~\ref{tlusty} we show a plot of the photospheric UV flux for a model
representing \zo\, near the \nspec{2}{3}{S}{1} \(\rightarrow\)
\nspec{2}{3}{P}{1,2} transitions of \ion{Mg}{11}. The model is taken from the
TLUSTY O star model grid \citep{LH03}. Note that for \ion{Mg}{11}, most of
the intercombination line strength still arises from the \nspec{2}{3}{P}{1} to
ground transition.

We also compute the \({\cal R}\) ratio using an averaged value of \(H_{\nu}\),
which we compare to the \({\cal R}\) ratio calculated using the non-averaged
(radially dependent) \(H_{\nu}\). We do this to understand whether it is
important to explicitly account for photospheric absorption lines, or whether
it is sufficient to calculate \({\cal R}\) using an averaged value of the
photospheric UV flux. We use the average value of \(H_{\nu}\) over the range
where \(0.1 < {\cal R} / {\cal R}_0 < 0.9\), or \(9 > 2\, P\, W(r) >
0.111\). There are two reasons for this: when the photoexcitation rate is much
less than the critical rate, the effect of photospheric lines on \(\cal R\) is
small; and when the photoexcitation rate is so high that the forbidden line is
very weak, we can't measure variations in the forbidden line strength.
We estimate this range using the continuum UV flux. In cases where \({\cal R}\)
does not ever get reduced to \(0.1 {\cal R}_0\) (even at the photosphere)
because the UV flux is not strong enough, we average from the rest frequency
to the frequency at which \({\cal R} = 0.9 {\cal R}_0\). 

In Figure~\ref{zetaorimgxi_tlusty} the dashed lines show \({\cal R}(r)\) for
averaged and non-averaged \(H_{\nu}\) for \ion{Mg}{11} for the star \zo. There
are substantial fluctuations in \({\cal R}(r)\) for the non-averaged case. The
solid lines in the figure are discussed in the following section; they
represent the effects of averaging the emission over a range of radii, as
opposed to simply over a range of frequencies.

In making this figure we have ignored all additional Doppler shifts, as the
purpose of the plot is mainly to illustrate qualitatively the effect of 
photospheric absorption lines on the \({\cal R}\) ratio. Examples of
potentially relevant Doppler shifts are the thermal velocities of the ions (of
order \(100\, {\mathrm{km\, s^{-1}}}\) for neon at 0.4 keV), stellar rotation
(typically \(100-200\, {\mathrm{km\, s^{-1}}}\) for O-type stars, although the
wind also rotates), and the non-monotonicity of the stellar wind due to shocks
\citep[e.g.][of order a few \(100\,{\mathrm{km\, s^{-1}}}\)]{F95}. We have
also treated the star as a point source rather than a finite disk, which would
change the projected velocity as a function of position on the stellar disk.
All of these effects are small compared to the wind terminal velocity, but
they could diminish the impact of photospheric lines on the $f/i$ ratio by
smearing out the photospheric spectrum.

One possibly important effect we neglect is scattering by resonance lines of
ions in the wind. This is probably relevant only for \ion{Mg}{11}. In this
case the \ion{O}{6} line at 1031.91 \AA\, is on the blue side of the
\nspec{2}{3}{S}{1} \(\rightarrow\) \nspec{2}{3}{P}{1} transition at 1034.31
\AA, which means that it could scatter the UV light from the photosphere to a
different wavelength. However, it is not clear that this will greatly affect
the line ratio, as the scattering process does not generally destroy
photons. The detailed effects of scattering by this transition could be
assessed by modelling the radiative transfer in the wind at this wavelength
range, but this is beyond the scope of this work.

\subsection{The integrated ratio \label{sub:Rbar}}

In the preceding two subsections we calculated the radial dependence of the
$f/i$ ratio. Here we will calculate the $f/i$ ratio integrated over an
emitting volume that may span a wide range of radii. After all, for any
realistic model of a stellar wind, we expect the X-ray emitting plasma to be
distributed over a large range of radii (although it could be a small range of
radii for a coronal model). We cannot directly observe the ratio as a function
of radius, but only the overall ratio, or the ratio as a function of the
observed Doppler shift.

We make the simple assumptions that the emissivity of the X-ray emitting
plasma scales as the wind density squared above some onset radius. This is the
same set of assumptions as the model of \citet{OC01}, with the two additional
simplifications that there is no continuum absorption and that there is no
radial variation in the X-ray filling factor.  These approximations are not
unreasonable, considering the low characteristic optical depths and the radial
dependence of the filling factor reported by  \citet{KCO03} for fits to line
profiles in the {\it{Chandra}} HETGS spectrum of \zp, especially for high $Z$,
where the optical depths are expected to be smallest.

To calculate the integrated strength of the forbidden and
intercombination lines, we weight the integrand with the normalized (radially
dependent) strength of each line.\footnote{We could instead express the
  integrated ratio as a single volume integral of the f-to-i ratio with a
  weighting term for the overall emissivity of the complex, but we feel that
  formalism we use here, of a ratio of two separate emissivity integrals, is
  more intuitive.  However, the two methods are formally equivalent.} The
weights are 
\begin{equation}
f(u) = G \frac{{\cal R}(u)}{1 + {\cal R}(u)}
\label{eqn:fu}
\end{equation}
and 
\begin{equation}
i(u) = G\frac{1}{1 + {\cal R}(u)}.
\label{eqn:iu}
\end{equation}
Here \(u \equiv R_* / r\) is the inverse radial coordinate. We have introduced
\(G \equiv (for + int) / res\) to ensure that the weighting factors are properly
normalized relative to the resonance line; we will discuss this in more
detail in the next section. The radial dependence of \({\cal R}(u)\) was
discussed in the previous sections (cf. Equation \ref{eqn:rofr}).

The integrated ratio is then
\begin{equation}
  \overline{\cal R}(u_0) = \frac{\int dV\, \eta_f}{\int dV\, \eta_i}
\end{equation}
where \(\eta_{f,i}\) are the emissivities of the forbidden and
intercombination lines. The integrals are
\begin{equation}
  \int dV\, \eta_{f,i} \propto \int^\infty_{R_0} \Omega(r) r^2 dr \rho^2(r)
  f,i(r) \propto \int_0^{u_0} \frac{du\, \Omega(u)}{w^2(u)} f,i(u)
\end{equation}
where we have used \(\rho(u) \propto u^2 / w(u)\). \(\Omega(u) = 2 \pi (1 +
\sqrt{1 - u^2})\) is the solid angle visible by the observer (i.e. not
obscured by the stellar core). \(w(u) = v(u) / v_{\infty} = (1 - u)^{\beta}\)
is the scaled velocity; we take \(\beta = 1\) as a convenient approximation,
as discussed in the following section. \(R_0\) is the onset radius for X-ray
emission, and \(u_0 = R_* / R_0\) is its inverse. The He-like line strength
weights \(f,i(u) \) are given by Equations~\ref{eqn:fu} and \ref{eqn:iu},
respectively.

\begin{figure}[p]
  \begin{center}
    \plotone{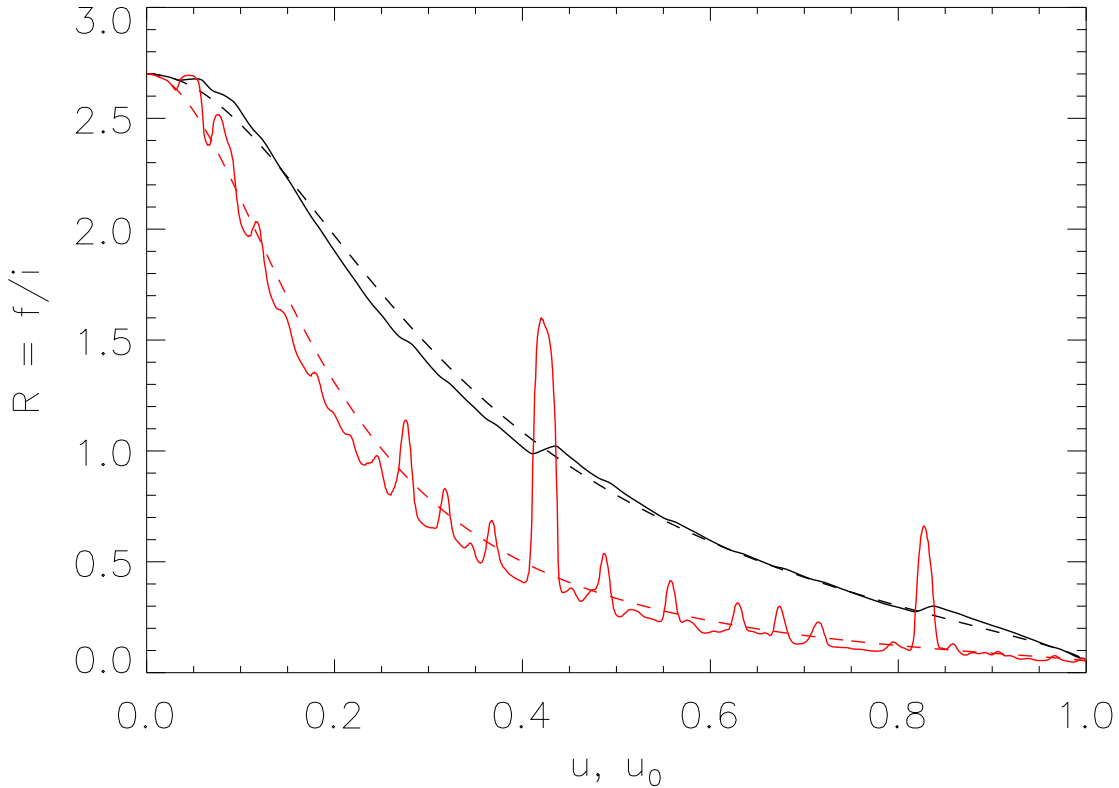}
  \end{center}
  \caption{The $f/i$ ratio for the \ion{Mg}{11} triplet of \zo\, plotted as a
    function of the inverse radial coordinate \(u=R_*/r\). The solid lines are
    for the actual model photospheric UV flux, while the dashed lines are for
    an averaged value. The red lines denote the local radial dependence of
    \({\cal R} (u)\) and the black lines show the integrated ratio
    \(\overline{\cal R} (u_0)\) observed for the whole star (see text). Note
    that $u$ and \(u_0\) are not comparable physical quantities, since $u$
    corresponds to a single radius, which could be interpreted as a
    characteristic radius, while \(u_0\) corresponds to the {\it minimum}
    radius for the onset of X-ray emission. The solid lines include the
    effects of the photospheric UV flux for transitions to both the
    \nspec{2}{3}{P}{1} and \nspec{2}{3}{P}{2} states, although the
    \nspec{2}{3}{P}{1} state is far more important for \ion{Mg}{11}. The peaks
    in the red solid line correspond to the absorption lines in the top panel
    of Figure~\ref{tlusty}.}
\label{zetaorimgxi_tlusty}
\end{figure}

In Figure~\ref{zetaorimgxi_tlusty}, the solid lines show the integrated $f/i$
ratio as a function of \(u_0\) for \ion{Mg}{11} in \zo. The
integrated ratio is very similar for both the averaged flux (black solid line)
and unaveraged (red solid line) flux cases. Since it is much simpler to
consider only a single value of photospheric UV flux and because it agrees
well with the more detailed treatment, we do so in the rest of
this paper. However, it should be noted that if one modeled the X-ray
emission as arising near a single radius or Doppler shift, as might be
appropriate for a coronal model, the actual photospheric flux (including
absorption lines) would have to be included in the modeling. 

It is important to note that there are two separate physical effects being
considered here: the first is the effect of using the actual photospheric
spectrum instead of a wavelength average, and the second is the averaging of
the \(\cal R\) ratio over a range
of radii. What Figure~\ref{zetaorimgxi_tlusty} shows is that the first effect
is not important if we include the second. However, when comparing the radius
inferred from a localized model to the {\it minimum} radius inferred from the
distributed model, it is crucial to realize that they are physically different
quantites. The radius in a localized model can be taken literally as the
characteristic location of the X-ray emitting plasma, but in the distributed
model, the minimum radius is the smallest radius where there is X-ray
emission; it can be interpreted physically as the shock onset radius.

\begin{figure}[p]
  \begin{center}
    \plotone{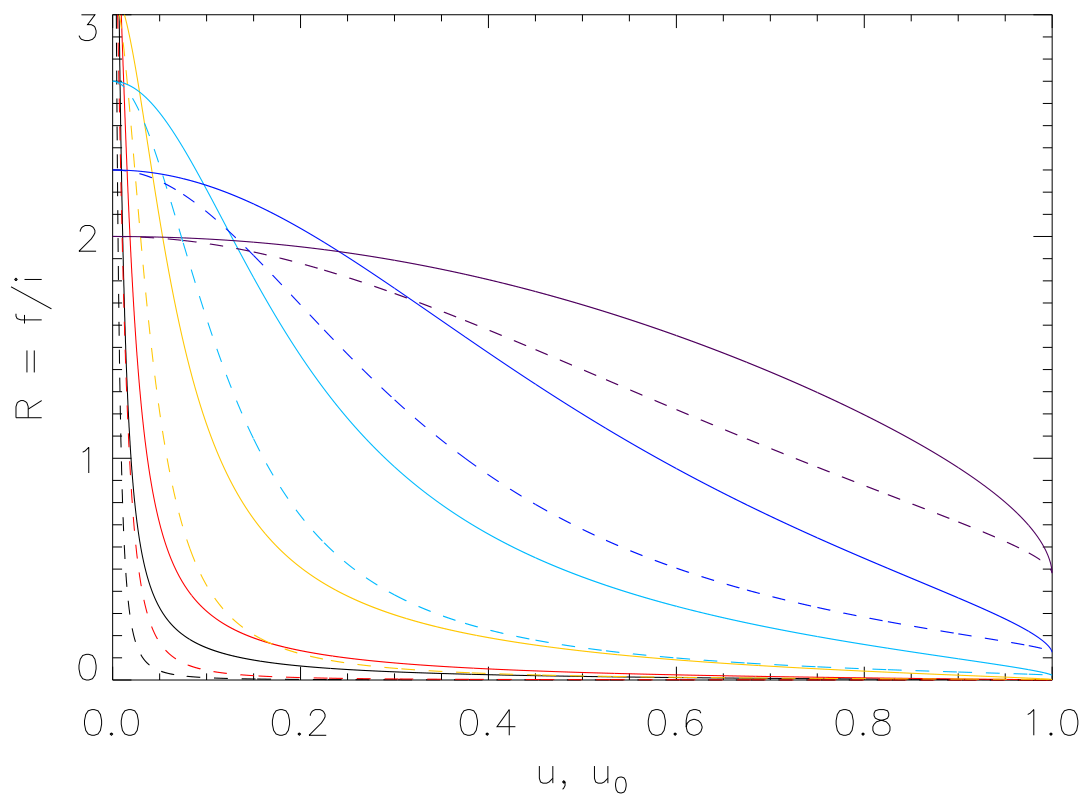}
  \end{center}
  \caption{The $f/i$ ratio for six He-like triplets observed in \zp. The
    dashed lines show the radial dependence of \({\cal R}\), while the solid
    lines show the dependence of the integrated ratio \(\overline{\cal R}\) on
    the inverse minimum radius \(u_0 = R_* / R_0\). The colors are black for
    \ion{N}{6}, red for \ion{O}{7}, orange for \ion{Ne}{9}, green for
    \ion{Mg}{11}, blue for \ion{Si}{13}, and purple for \ion{S}{15}.}
\label{zetapuphelike}
\end{figure}

\begin{figure}[p]
  \begin{center}
    \plotone{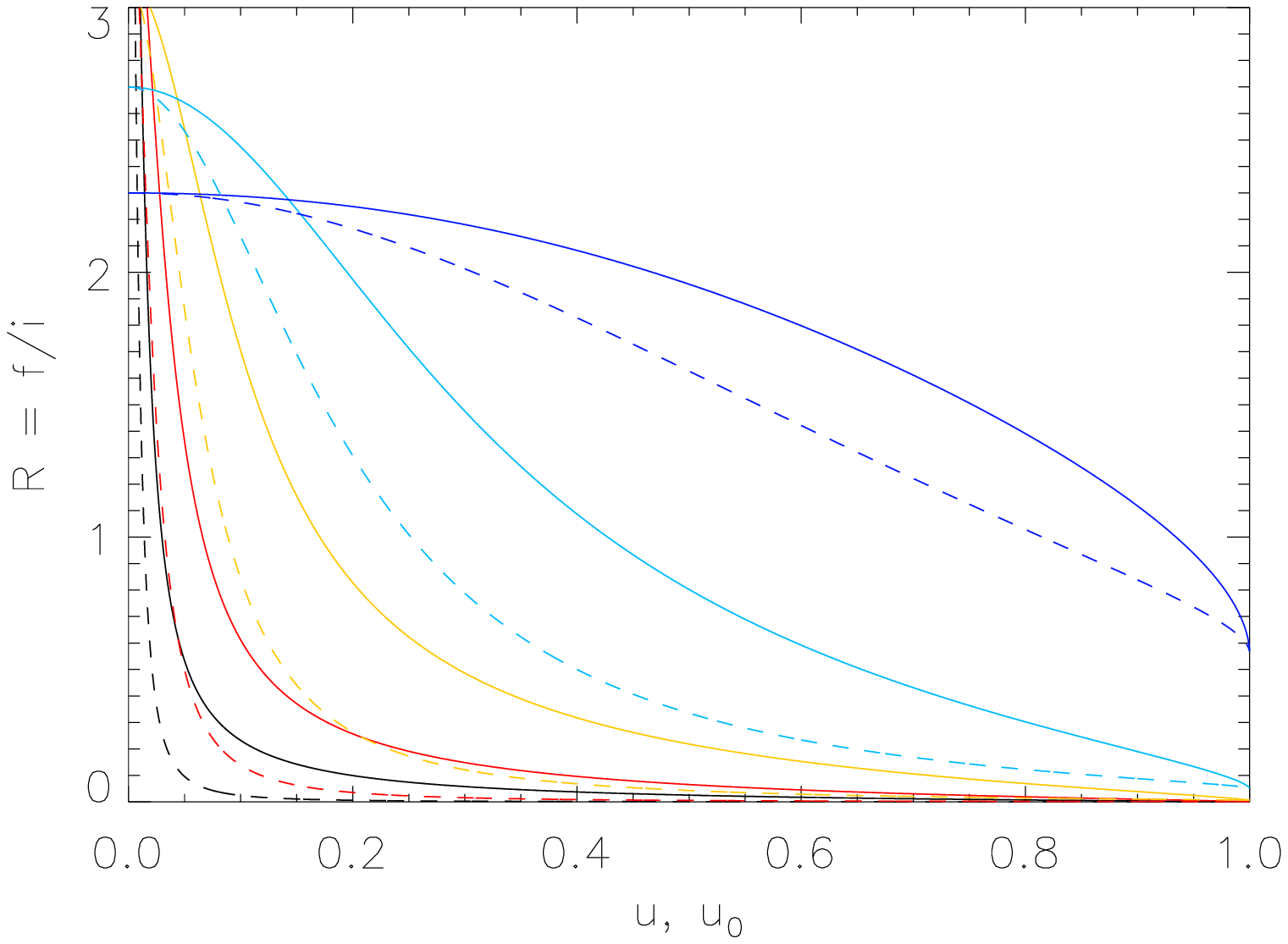}
  \end{center}
  \caption{The $f/i$ ratio for five He-like triplets observed in \zo. Scheme is
    as in Figure~\ref{zetapuphelike}.}
\label{zetaorihelike}
\end{figure}

\begin{figure}[p]
  \begin{center}
    \plotone{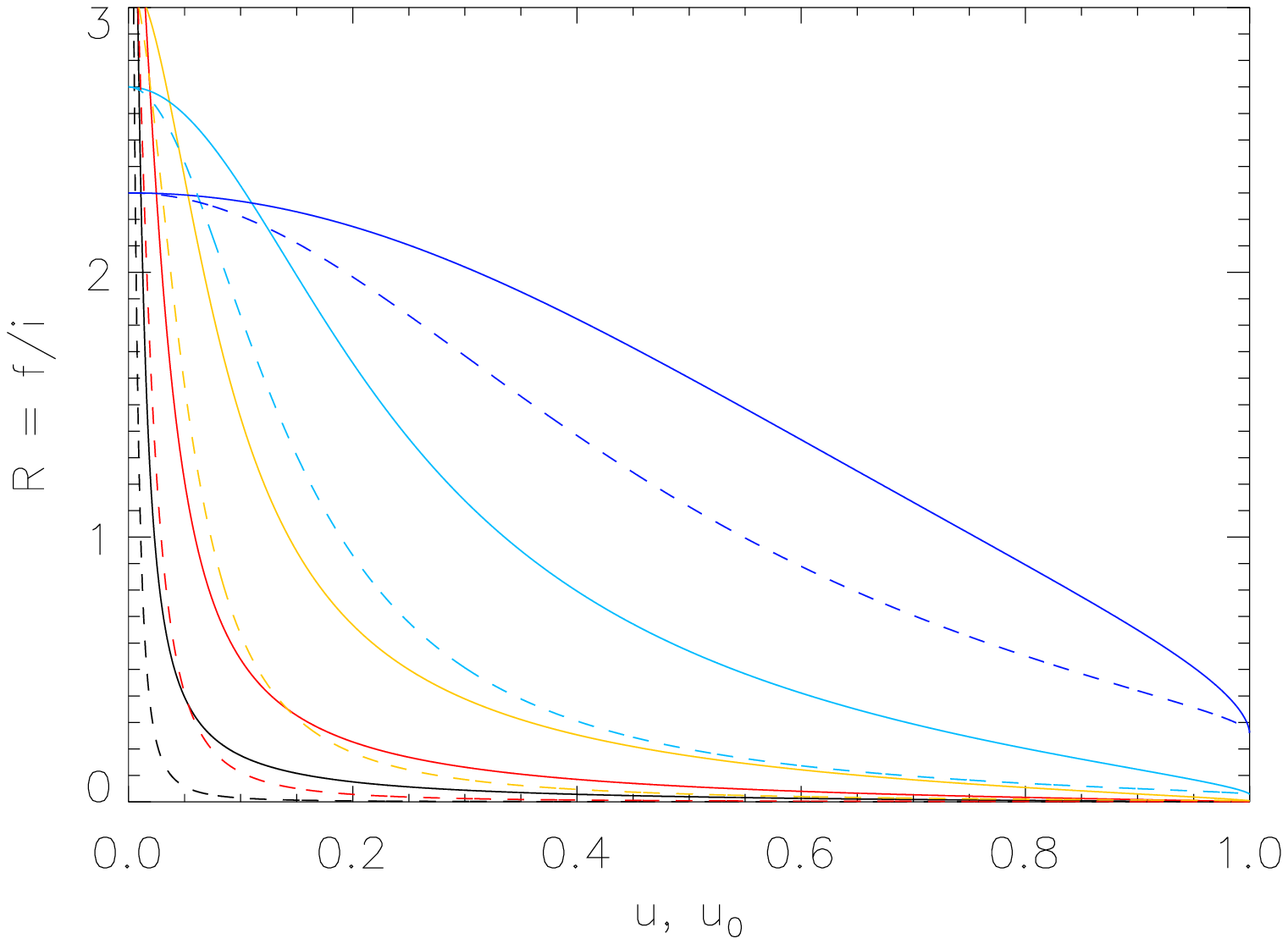}
  \end{center}
  \caption{The $f/i$ ratio for five He-like triplets observed in \io. Scheme is
    as in Figure~\ref{zetapuphelike}.}
\label{iotaorihelike}
\end{figure}

\begin{figure}[p]
  \begin{center}
    \plotone{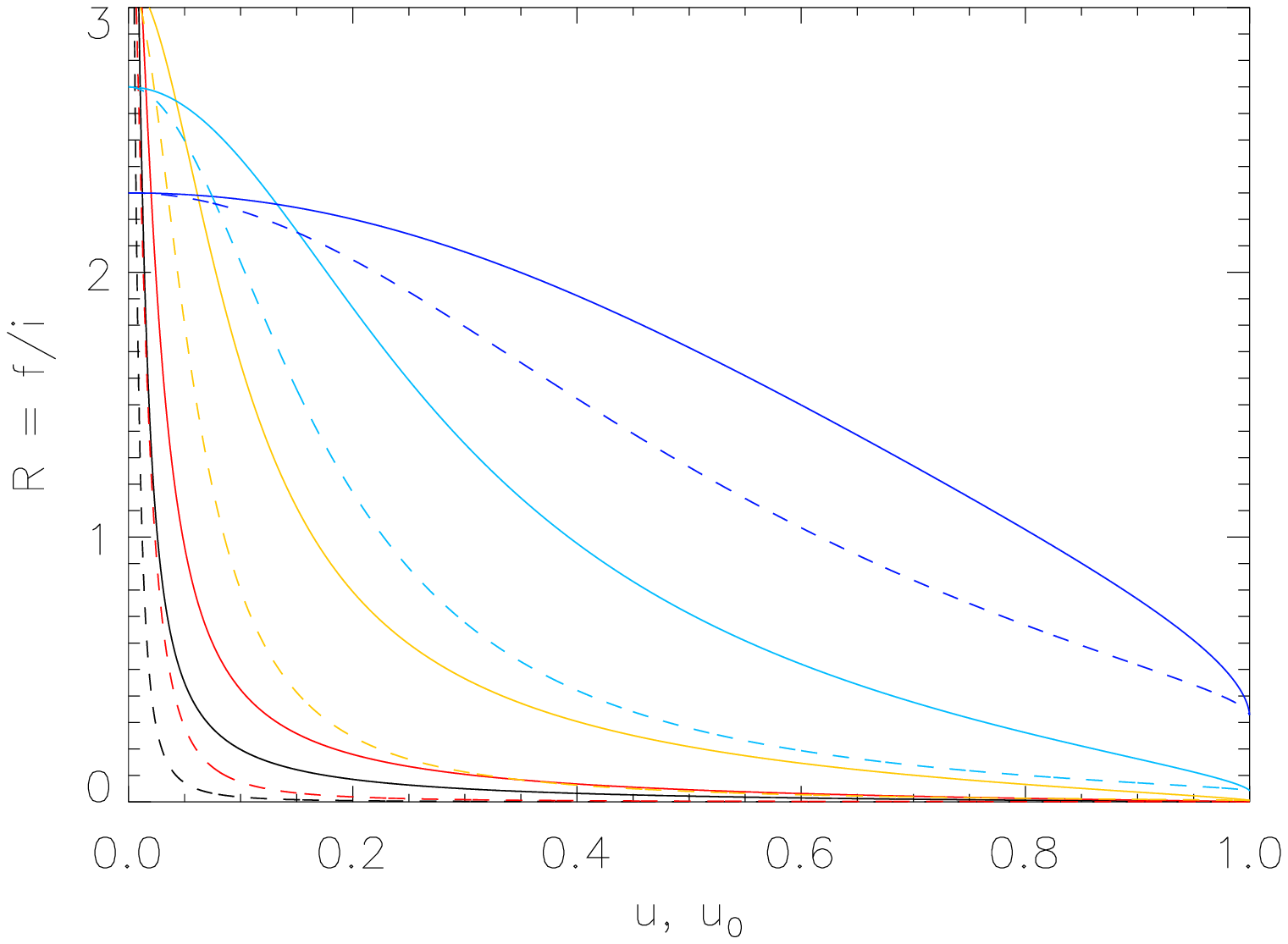}
  \end{center}
  \caption{The $f/i$ ratio for five He-like triplets observed in \deltao. Scheme
    is as in Figure~\ref{zetapuphelike}.}
\label{deltaorihelike}
\end{figure}

In Figures~\ref{zetapuphelike},\ref{zetaorihelike},\ref{iotaorihelike}, and
\ref{deltaorihelike}, we show \({\cal R}(u)\) and \(\overline{\cal R}(u_0)\)
for all He-like ions observed in the four O stars we consider in this
paper. These plots all assume an averaged value of the photospheric UV
flux. For a given measured value of \({\cal R}\), there are substantial
differences between the value of \(u_0\) derived assuming a distributed plasma
and the value of $u$ derived assuming a plasma dominated by one radius - that
is, \(u_0\) is always larger than $u$ for a single radius, as one would
expect. 

In Table~\ref{TabCompareTLUSTYKurucz} we compare our calculations using TLUSTY
model stellar atmosphere fluxes to the same calculations using \cite{K79}
fluxes, as in \citet{WC01} and \citet{CMWMC01}. We make the comparison for one
key ion for each paper, both of which have their \nspec{2}{3}{S}{1}
\(\rightarrow\) \nspec{2}{3}{P}{J} transition wavelengths in the Lyman
continuum. We use \({\cal R}_0\) values taken directly from the plots of
\citet{WC01} and \citet{CMWMC01}. For most ions in these two papers, \({\cal
  R}_0\) is taken from BDT, but for \ion{Si}{13}, \citet{WC01} use \({\cal
  R}_0 = 2.85\), while the BDT value is 2.51. The values of \({\cal R}_0\)
given in BDT are systematically higher than those in \citet{PMDRK01}.

There are substantial differences between our calculations of \({\cal
  R}_{ph}\) (the value of \({\cal R}\) at the photosphere) and those of
\citet{WC01}, \citet{CMWMC01}, and \citet{MCWMC02}. These differences mainly
arise from differences in the continuum flux of the photospheric models
shortward of the Lyman edge; the TLUSTY models generally predict a factor of
2-3 more than the Kurucz models. 

The combination of the different Lyman continua and \({\cal R}_0\) values lead
to substantially higher values of \({\cal R}_{ph}\) for \ion{Si}{13} and
\ion{S}{15} in \citet{WC01}, \citet{CMWMC01}, and \citet{MCWMC02}. This means
that we would infer systematically larger radii than these authors, given the
same measured value of \({\cal R}\).

Regardless of the differences between TLUSTY and Kurucz model atmospheres,
there are substantial uncertainties in the Lyman flux of any model atmosphere;
this part of the spectrum is generally inaccessible to observation, and the
models' Lyman continua have not been directly verified experimentally. 
In the two cases where early B stars have been directly observed in the Lyman
continuum with {\it EUVE}, the fluxes have been roughly an order of 
magnitude above models \citep{Cel95, Cel96}; however, it should be pointed out
that these stars are significantly cooler than the O stars we are studying, so
that their Lyman fluxes are more sensitive to changes in the temperature
structure in the outer atmosphere. Furthermore, the
effective temperature scale used for O stars in the past may be systematically
too high \citep*{MSH02}, which would also have more of an effect on the part
of the spectrum shortward of the Lyman break. However, the effect of the
uncertainty in the model Lyman continuum flux is significantly larger than the
effect of the correction to the effective temperature scale.

\subsection{He-like line profiles \label{sub:hewind}}

Although it may sometimes be easier to measure the $f/i$ ratio directly and
compare it to a calculation for the ratio as a function of distance from the
star, it is potentially much more powerful to calculate line profiles
including the radial dependence of the line ratio and compare these to the
data. The expression for the line profile derived in OC is
\begin{equation}
L_x = C \int_0^{u_x} du \frac{f_X(u)}{w^3(u)}e^{-\tau(u, x)}
\end{equation}
In this expression, the volume filling factor of X-ray emitting plasma is
\(f_X(u)\propto u^q\), while $x$ refers to the velocity-scaled
dimensionless Doppler-shift parameter. \(\tau(u, x)\) is the optical depth
along the line of sight to the observer, which is usually written as the
product of a geometrical integral, \(t(u,x)\), and a dimensionless constant,
\(\tau_* = \frac{\kappa \dot{M}}{4 \pi v_{\infty} R_*}\), the characteristic
optical depth. It should be noted that the expression for the optical depth is
only analytic for integral values of the velocity law index \(\beta\);
otherwise it must be evaluated numerically. Because the expression for \(L_x\)
must also be evaluated numerically, it is preferable to take \(\beta\) to be
an integer in order to avoid a multidimensional integral. \(\beta=1\) is the
best integer approximation for most O stars (see, e.g., \citet{Pel06} for
models that include clumping).

To account for the relative line strengths of the triplet, we simply multiply
the integrand with  the weighting factors \(f(u) = G \frac{{\cal R}(u)}{1 +
  {\cal R}(u)}\) or \(i(u) = G \frac{1}{1 + {\cal R}(u)}\). This normalizes
the forbidden and intercombination lines to the resonance line, which may be
calculated using the above expression with no modification. If it is desirable
to normalize the sum of all three weighting factors to unity, one may divide
them by \(1+G\). In this work we have assumed that \(G\) does not vary with
radius. Although \(G\) does depend on temperature, the variation is not
strong, and the X-ray emitting plasma is likely multiphase. If there
is any variation in the line profile shapes caused by a radial dependence in
\(G\), it is not likely to be detectable except with data of very high
statistical quality.

In comparison with the integrated plots presented in the previous section, a
line profile with \(\tau_* > 0\) has a higher \({\cal R}\) ratio than one
with no absorption, given the same value of \(u_0\). This is
because the forbidden line is only formed farther out where absorption is
less, while the intercombination line is mainly formed close to the star, where
absorption is greater. Nonzero positive values of $q$ cause \({\cal R}\) to go
down, because relatively more emission comes from close to the star, while
negative values cause \({\cal R}\) to go up. 

In comparison with normal line profiles, the intercombination line has weaker
wings, as it becomes much weaker far away from the star. On the other
hand, the forbidden line is relatively flat topped; because of
photoexcitation, the profile appears as if it has a larger
effective value of \(R_0\) than the resonance line. 

The addition of the radial dependence of $f/i$ ratio to the OC profile model 
has the appealing property of enforcing self-consistency between the
radial dependences of the Doppler profile and the $f/i$ ratio. Also, although
it does make the quite reasonable assumption that the X-ray emitting plasma
follows the same \(\beta\)-velocity law as the wind, it is not tied to any
particular heating mechanism.

In the next section, we use this model to fit Chandra HETGS spectra of
four O stars.

\section{Data reduction and analysis \label{sec:data}}

In this section we fit He-like triplets in the {\it Chandra} HETGS data of
four O stars: \zp, \zo, \io, and \deltao. We only fit \ion{Mg}{11},
\ion{Si}{13}, and for \zp\, \ion{S}{15}. This is because \ion{Ne}{9} and lower
$Z$ He-like species generally have \({\cal R} < 0.2\) in O stars and therefore
do not contain significant information in the line ratio.

\subsection{Data processing}

Primary data products were obtained from the {\it Chandra} data archive and
processed using standard CIAO routines outlined in the CIAO grating
spectroscopy threads.\footnote{http://cxc.harvard.edu/ciao/threads/gspec.html}
The versions used were CIAO~3.1 and CALDB~2.28. The spectral fitting was done
with XSPEC~11.3.1. The $C$ statistic \citep{C79} is used instead of \(\chi^2\)
because of the low number of counts per bin. For \zo\, and \io\, the data were
split into two observations each, which were fit simultaneously. Emission
lines were fit over a wavelength range of (\(\lambda_r(1-v_{\infty}/c)-\Delta
\lambda\), \(\lambda_f(1+v_{\infty}/c)+\Delta \lambda\)), where \(\Delta
\lambda\) is the resolution of MEG at that wavelength. This range was chosen
to include the entire emission line, but at the same time to prevent the
quality of the continuum fit from influencing the fit statistic for the line. 
To get the continuum strength for a given line, we first fit it outside this
range, but near the wavelength of the line.

Because the MEG has substantially more effective area than the HEG at longer
wavelengths, we used only the MEG \(\pm 1\) order data for \ion{Si}{13} in
\zp\, and for \ion{Mg}{11} for all stars. For the \ion{S}{15} complex in \zp\,
and the \ion{Si}{13} complex in the other stars, the statistics are poorer, and
the contribution of the HEG is significant, so we simultaneously fit both the
HEG and MEG \(\pm 1\) order data. The MEG \(\pm 1\) order data for
\ion{Si}{13} in \zo\, are inconsistent, so we fit each of them seperately;
this inconsistency is discussed in more detail in the results subsection.

\subsection{Fitting procedure}

We use two different fitting procedures, depending on the number of counts in
the triplet. For triplets with many counts (\ion{Mg}{11} for all stars, and
\ion{Si}{13} for \zp), we fit them with the He-like OC profile described in
\S~\ref{sub:hewind}. The fixed model parameters are the line rest
wavelengths, the terminal velocity of the wind, the velocity law index
\(\beta=1\), the unaltered $f/i$ ratio \({\cal R}_0\), and the averaged
photospheric UV strength. The fit parameters from the profile model are $q$,
\(\tau_*\), and \(u_0\), in addition to the $G$ ratio and the overall
normalization. The four fit parameters other than normalization are fit on a
grid with spacing 0.2 for $q$ and \(\tau_*\), and spacing 0.05 for \(u_0\) and
$G$.

For lines with few counts (\ion{S}{15} for \zp\, and \ion{Si}{13} for
the other stars), we fit a three Gaussian model to prevent overinterpretation.
Rather than using three individual Gaussians, which would have three
separate normalizations, we use a model with parameters $G$, \({\cal R}\), the
overall normalization, and the velocity width, which is taken to be the same
for all the lines in a given complex. This avoids fitting problems due to
covariance in individual line normalizations, which can be a problem in
blended line complexes. It also allows us to directly measure the line ratios
and their errors, which are the quantities of interest. We fit the parameters
on a grid with spacing \(2\times 10^{-3}\) for \(\sigma_v\), 0.2 for \({\cal
  R}\), and 0.1 or 0.2 for $G$. We interpret the results of these
multi-Gaussian fits using the integrated ratio formalism described in
\S~\ref{sub:Rbar} and shown in
Figures~\ref{zetapuphelike}-\ref{deltaorihelike}.

In all cases we add a continuum component to approximate bremsstrahlung
emission. This is represented by a power law of index 2 with normalization
chosen to fit the continuum near the line. Care is taken to avoid including
moderately weak spectral lines in the continuum fit. A power law of index 2 is
not necessarily appropriate for the continuum in general, but over a
sufficiently short range in wavelength, any reasonable continuum shape is
statistically indistinguishable. An index of 2 is chosen because this gives a
flat continuum when \(F_{\lambda}\) is plotted versus wavelength.

We do not expect any other strong lines to contaminate our line
fits. \ion{Mg}{12} Ly $\gamma$ is at approximately the same wavelength as the
\ion{Si}{13} forbidden line, but even in \zo, where \ion{Si}{13} is
relatively weak, the strength of \ion{Mg}{12} Ly $\gamma$ expected based on
the strength of \ion{Mg}{12} Ly $\alpha$ is not enough to affect our
measurements significantly.


\section{Results \label{sec:results}}

\begin{figure}[p]
  \begin{center}
    \plotone{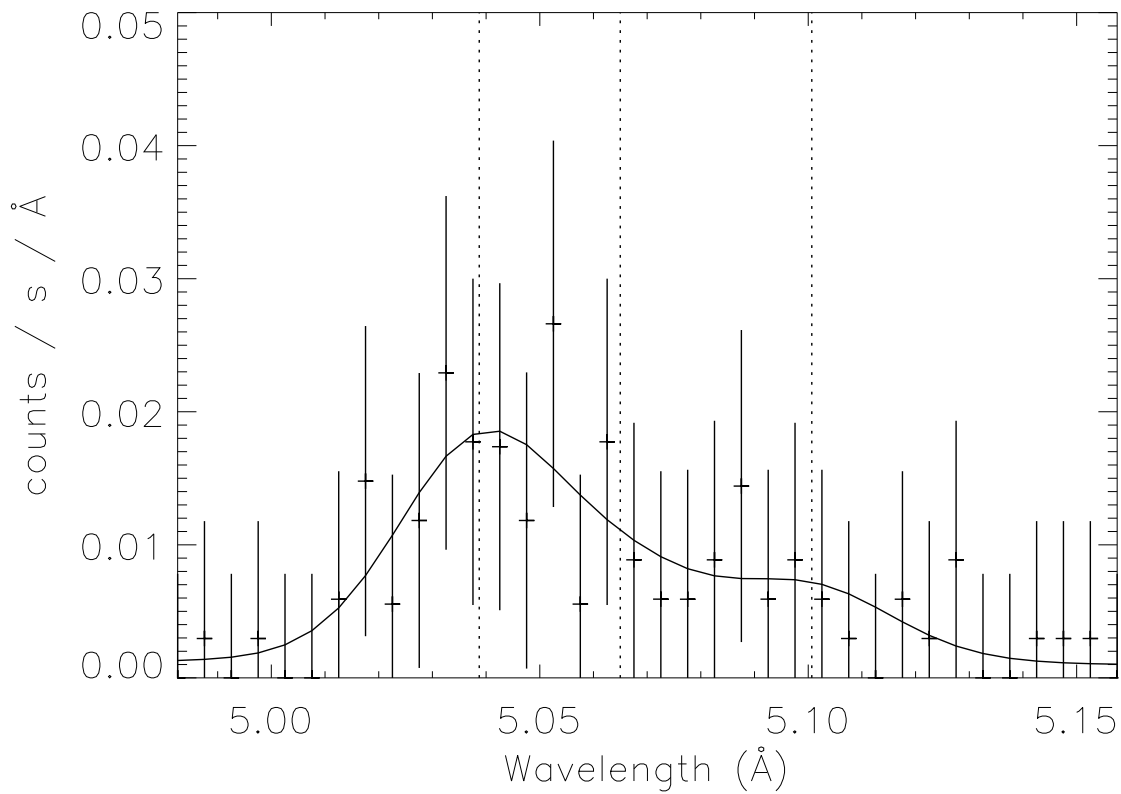}
  \end{center}
  \caption{MEG data and best-fit model for \ion{S}{15} in \zp. The positive
    and negative first order data have been coadded. The data are shown with
    error bars, and the model is shown as a solid line. The rest wavelengths
    of the resonance, intercombination, and forbidden lines are shown with
    dotted lines. This scheme is used in all subsequent figures presenting the
    data. Except where stated explicitly, the plots of Gaussian fits show the
    joint best fit to both the HEG and MEG data, even though data from only
    one grating are presented at a time.}
\label{ZetapupSXVMEG}
\end{figure}

\begin{figure}[p]
  \begin{center}
    \plotone{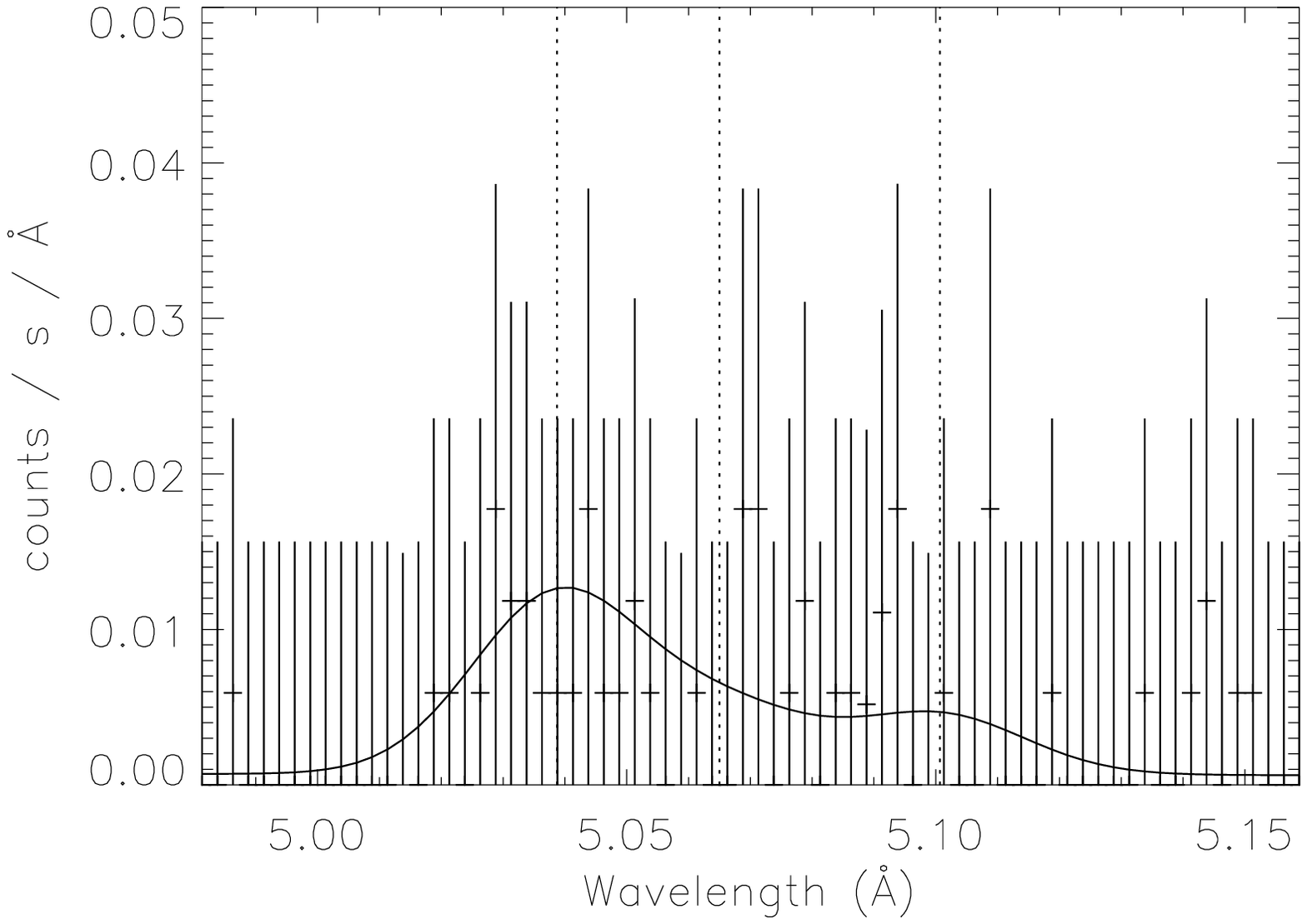}
  \end{center}
  \caption{HEG data and best-fit model for \ion{S}{15} in \zp. The positive
    and negative first order data have been coadded.}
  \label{ZetapupSXVHEG}
\end{figure}

\begin{figure}[p]
  \begin{center}
    \plotone{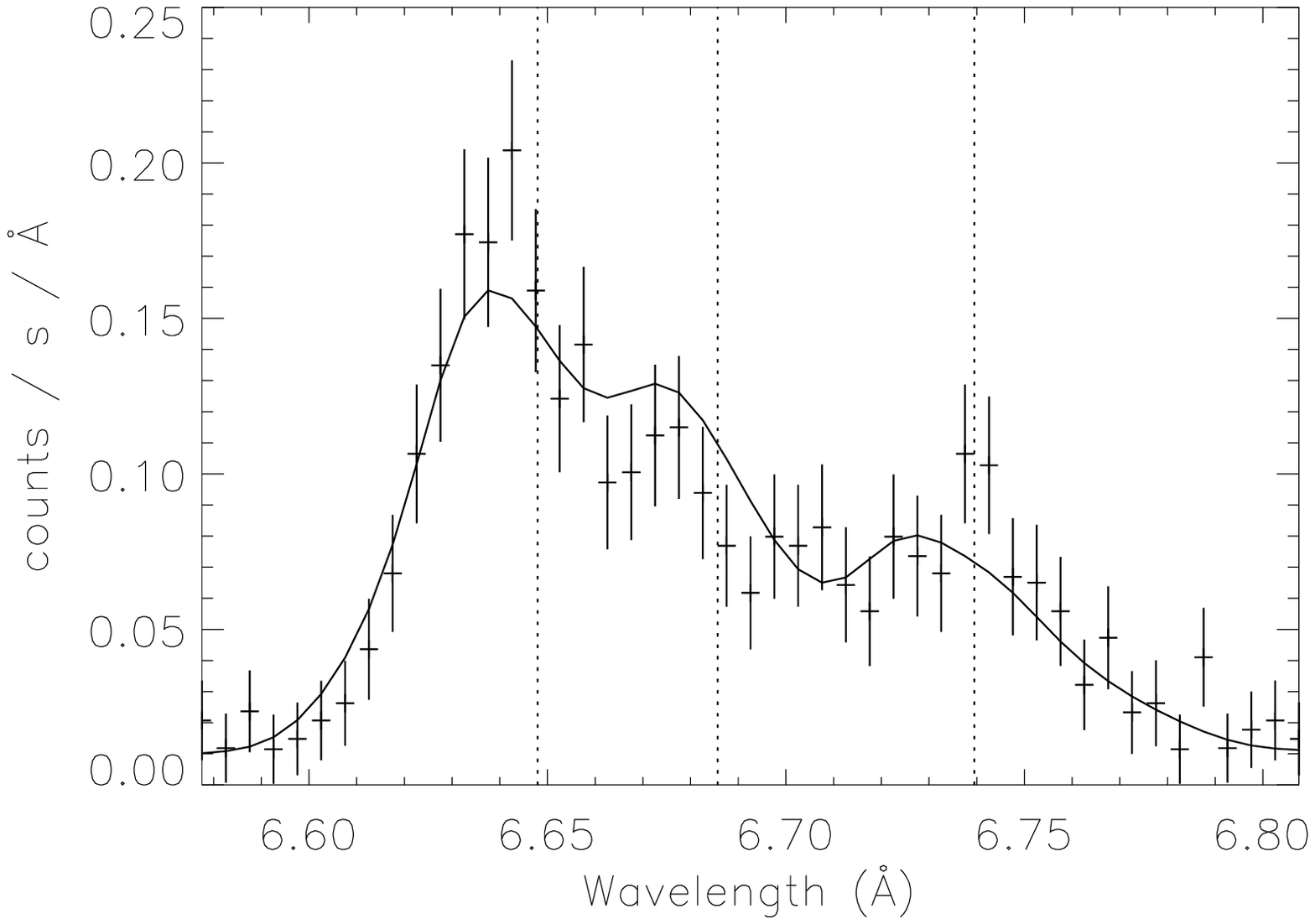}
  \end{center}
  \caption{MEG data and best-fit model for \ion{Si}{13} in \zp. The positive
    and negative first order data have been coadded.}
\label{ZetapupSiXIII}
\end{figure}

\begin{figure}[p]
  \begin{center}
    \plotone{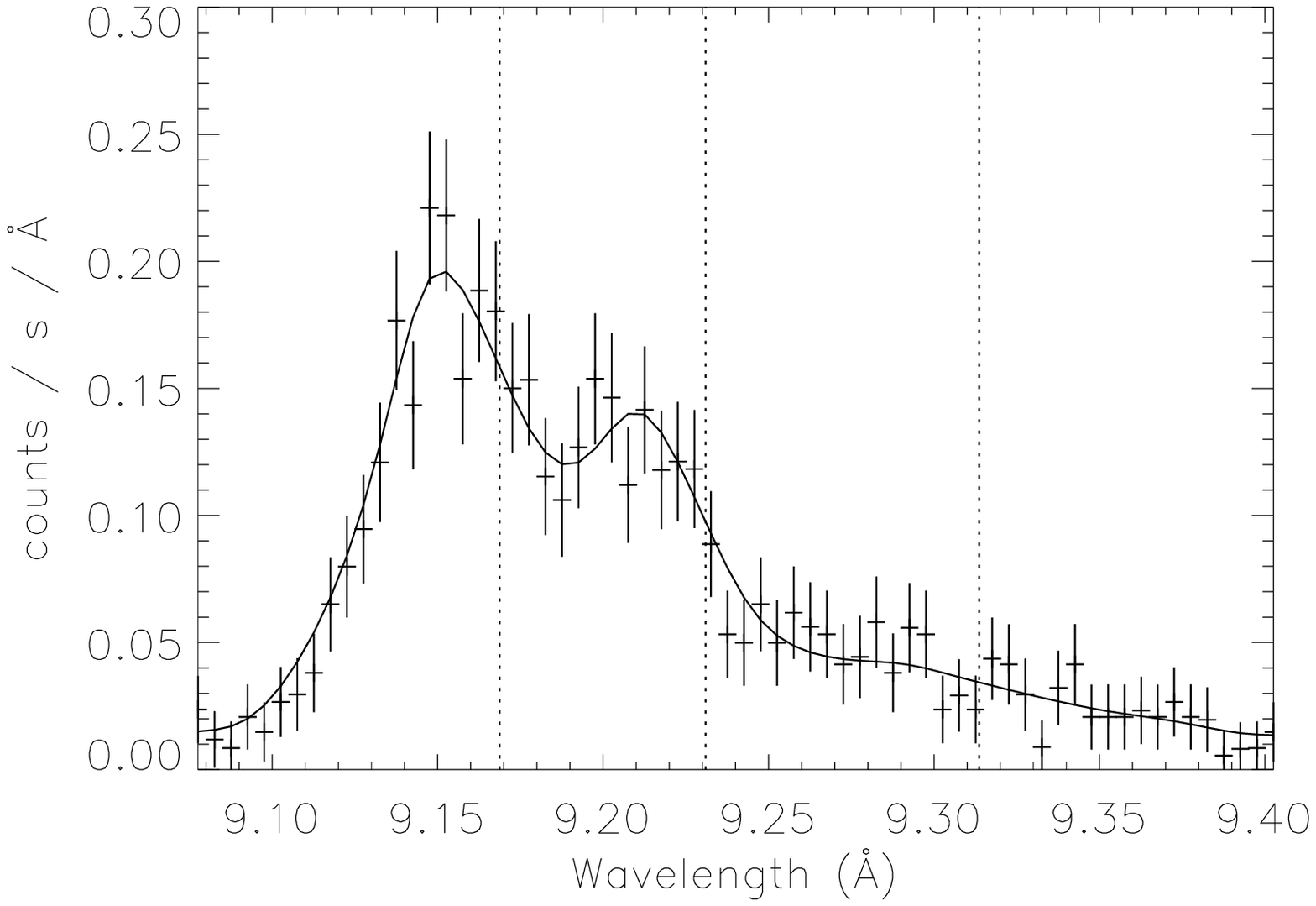}
  \end{center}
  \caption{MEG data and best-fit model for \ion{Mg}{11} in \zp. The positive
    and negative first order data have been coadded.}
\label{ZetapupMgXI}
\end{figure}

\begin{figure}[p]
  \begin{center}
    \plotone{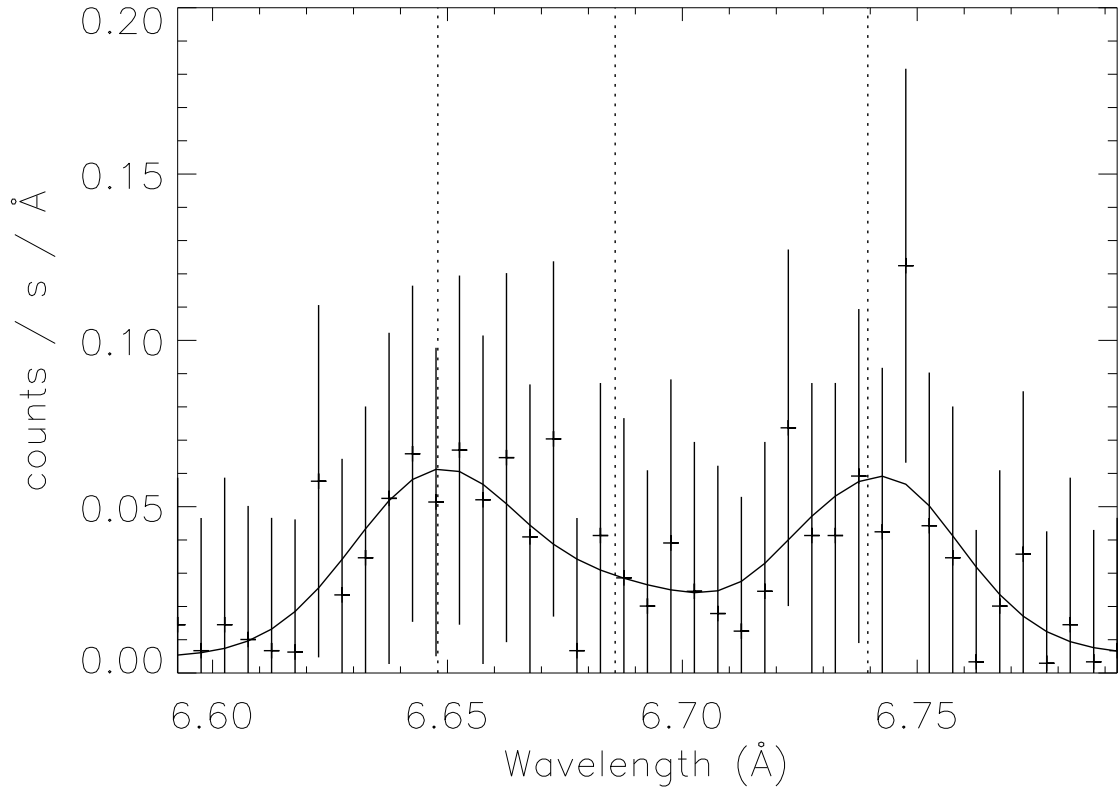}
  \end{center}
  \caption{MEG data and best-fit model for \ion{Si}{13} in \zo. The positive
    and negative first order data have been
    coadded. Figures~\ref{ZetaoriSiXIII+1} and \ref{ZetaoriSiXIII-1} show the
    positive and negative first order MEG data separately.} 
\label{ZetaoriSiXIIIMEG}
\end{figure}

\begin{figure}[p]
  \begin{center}
    \plotone{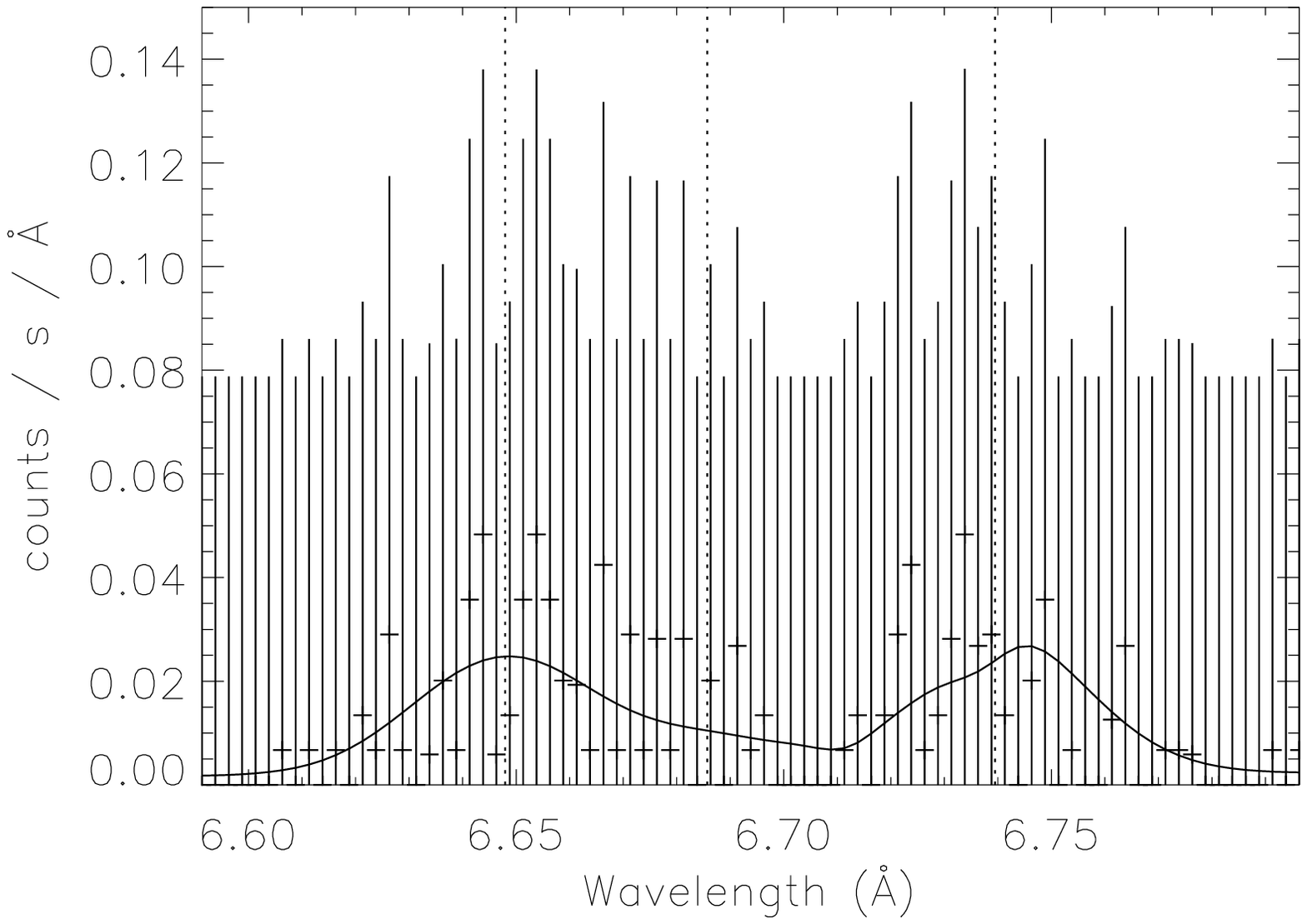}
  \end{center}
  \caption{HEG data and best-fit model for \ion{Si}{13} in \zo. The positive
    and negative first order data have been coadded.}
\label{ZetaoriSiXIIIHEG}
\end{figure}

\begin{figure}[p]
  \begin{center}
    \plotone{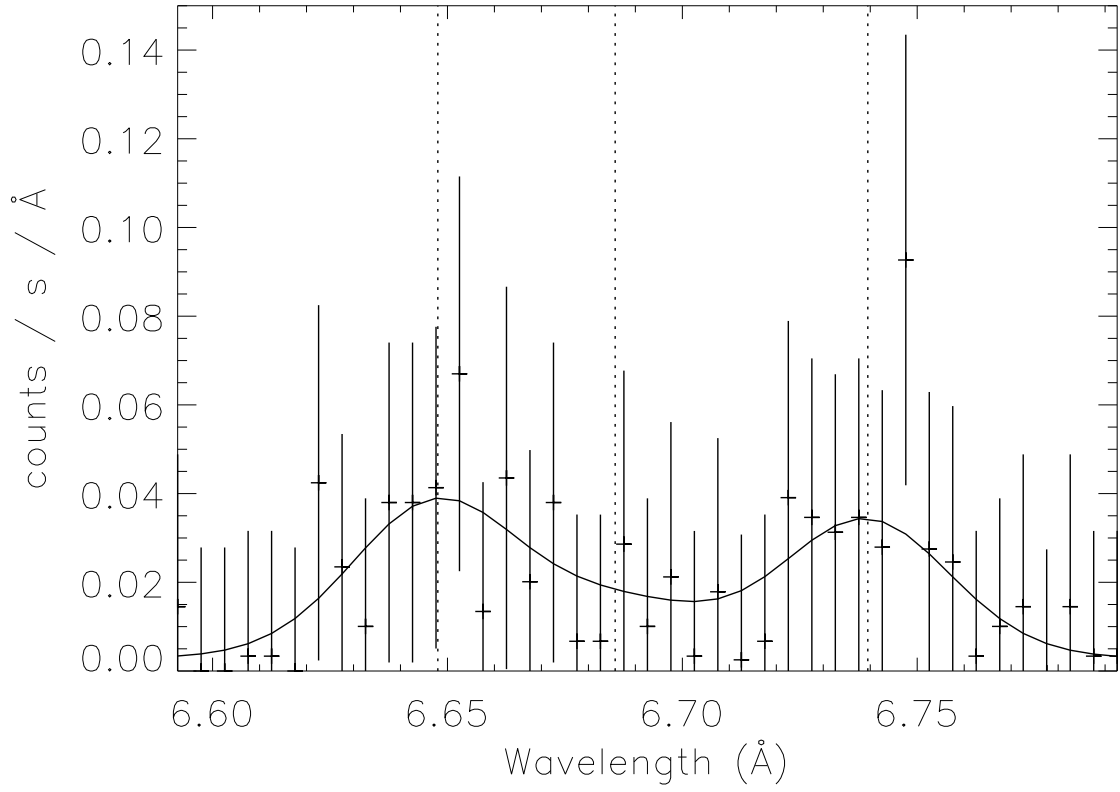}
  \end{center}
  \caption{MEG positive first order data and best-fit model for \ion{Si}{13}
    in \zo. The model is the best fit to {\it only} the positive first order
    data.}
\label{ZetaoriSiXIII+1}
\end{figure}

\begin{figure}[p]
  \begin{center}
    \plotone{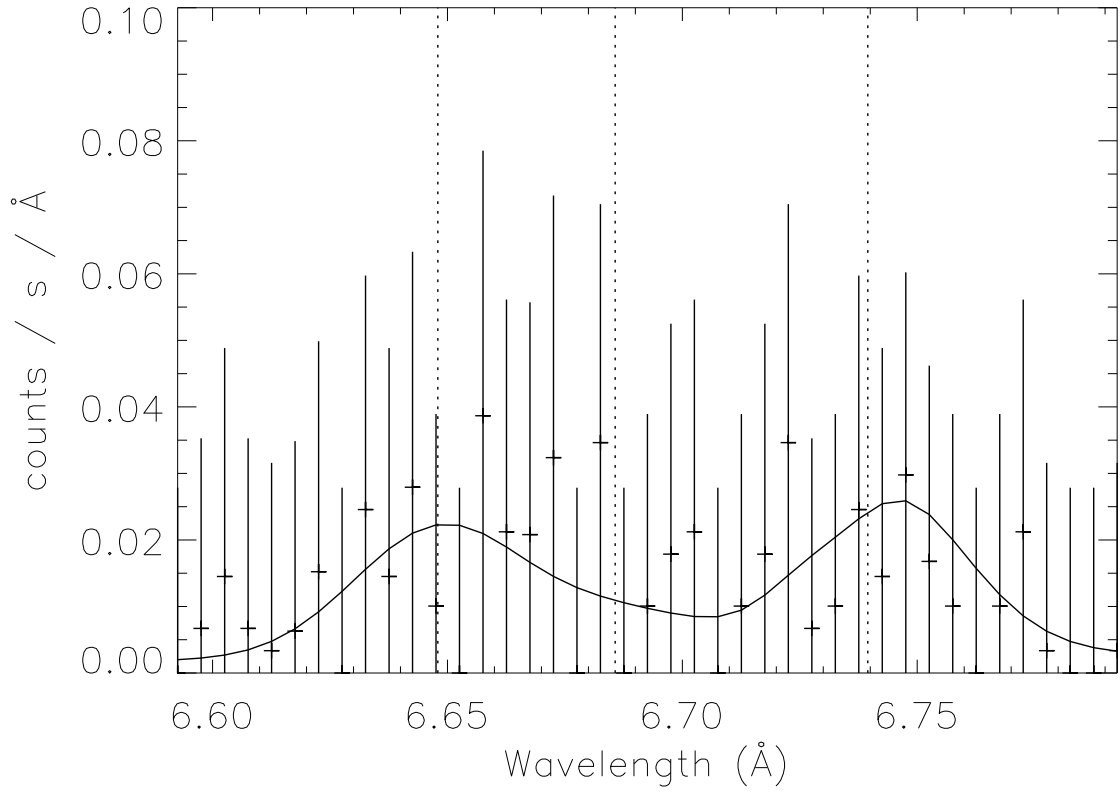}
  \end{center}
  \caption{MEG negative first order data and best-fit model for \ion{Si}{13}
    in \zo. The model is the best fit to {\it only} the negative first order
    data.}
\label{ZetaoriSiXIII-1}
\end{figure}

\begin{figure}[p]
  \begin{center}
    \plotone{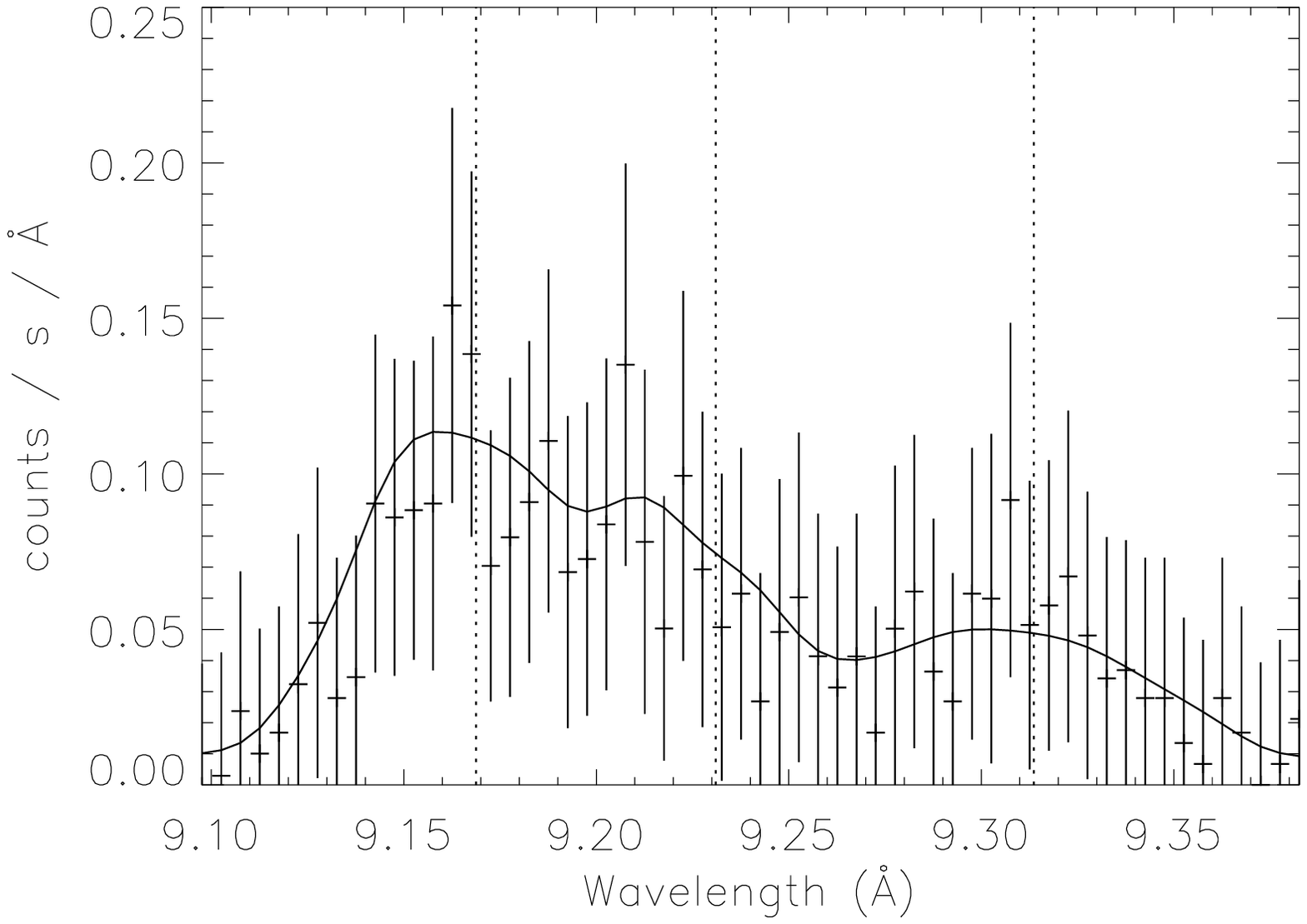}
  \end{center}
  \caption{MEG data and best-fit model for \ion{Mg}{11} in \zo. The positive
    and negative first order data have been coadded.}
\label{ZetaoriMgXI}
\end{figure}

\begin{figure}[p]
  \begin{center}
    \plotone{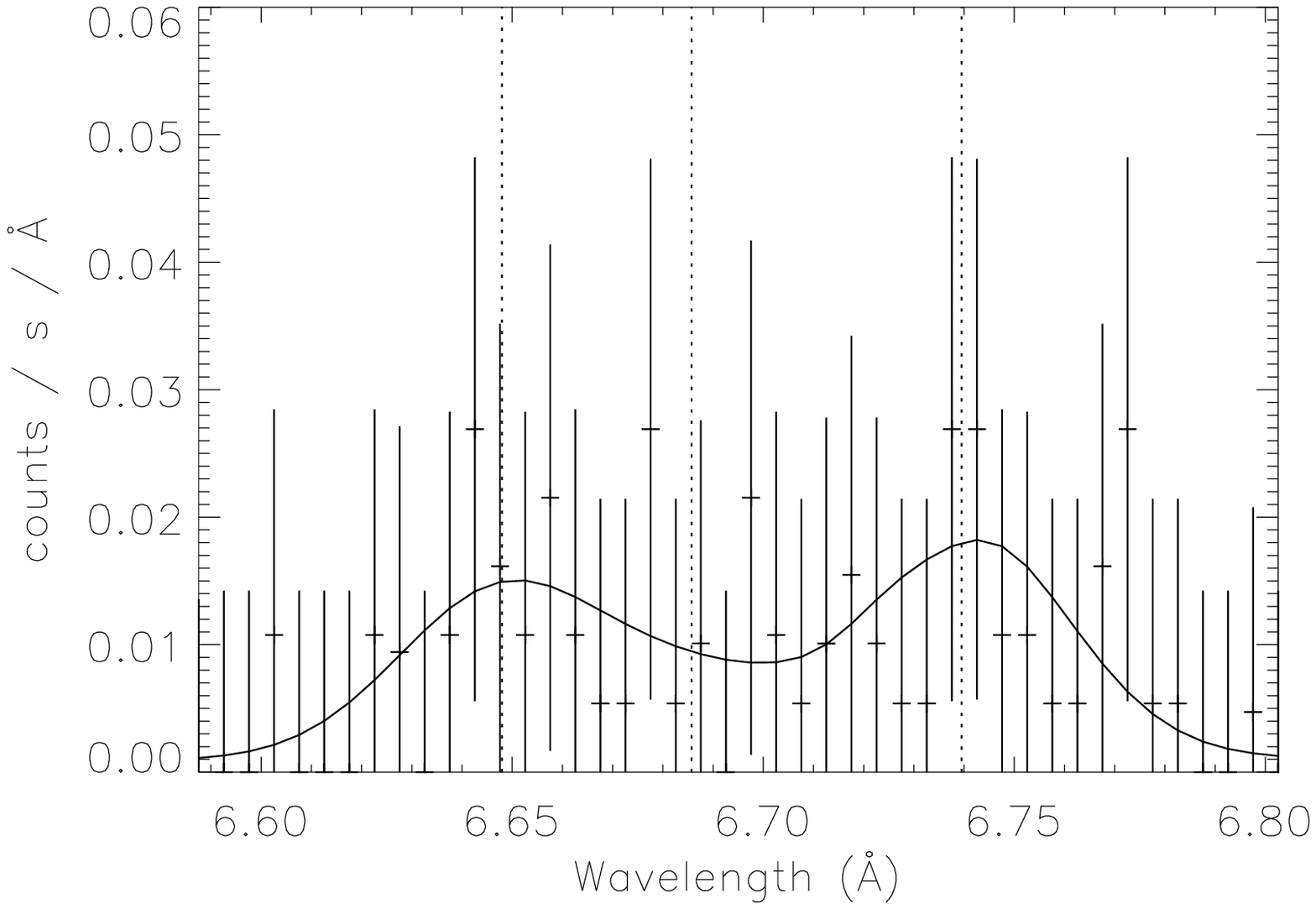}
  \end{center}
  \caption{MEG data and best-fit model for \ion{Si}{13} in \io. The positive
    and negative first order data have been coadded.}
\label{IotaoriSiXIIIMEG}
\end{figure}

\begin{figure}[p]
  \begin{center}
    \plotone{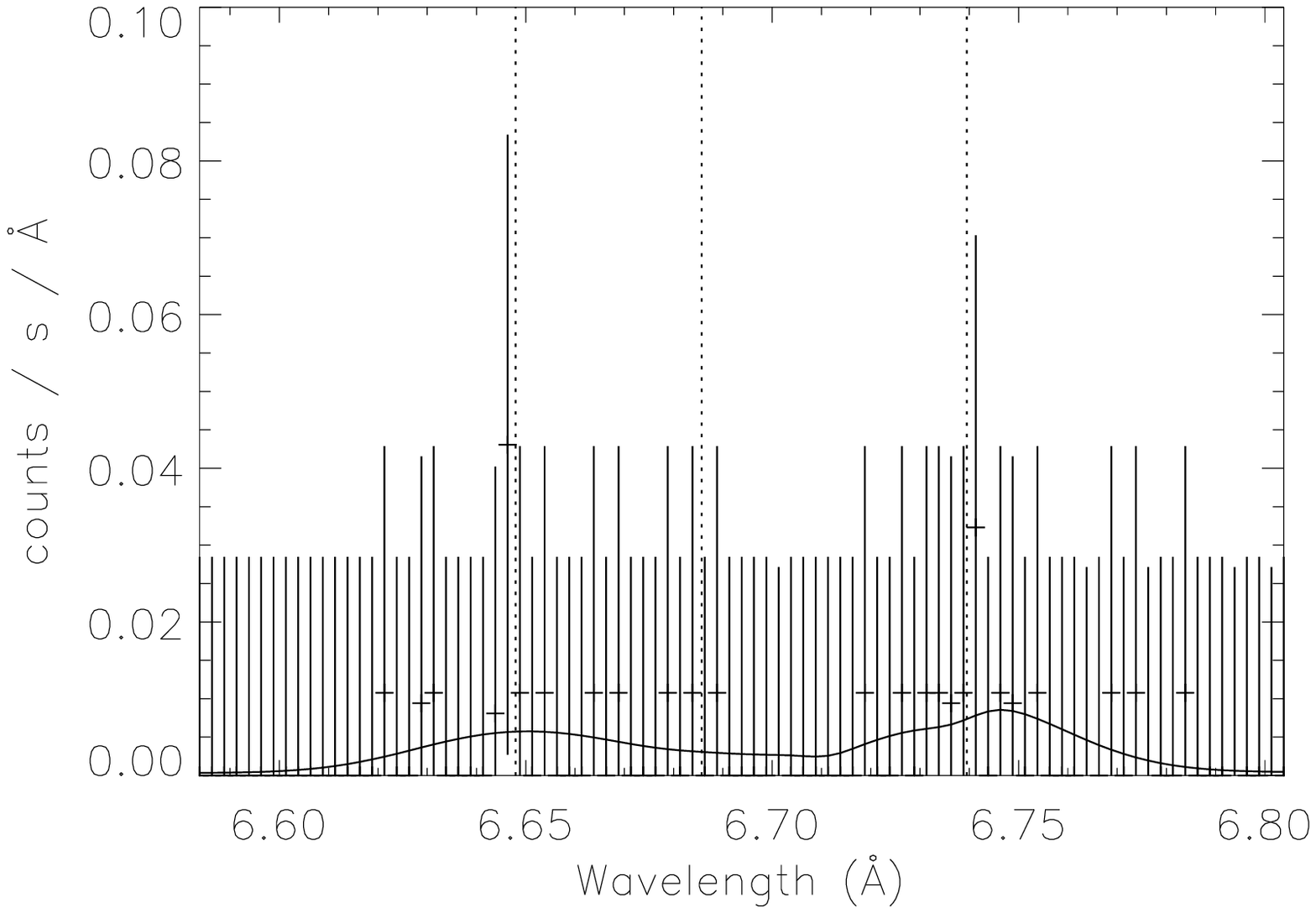}
  \end{center}
  \caption{HEG data and best-fit model for \ion{Si}{13} in \io. The positive
    and negative first order data have been coadded.}
\label{IotaoriSiXIIIHEG}
\end{figure}

\begin{figure}[p]
  \begin{center}
    \plotone{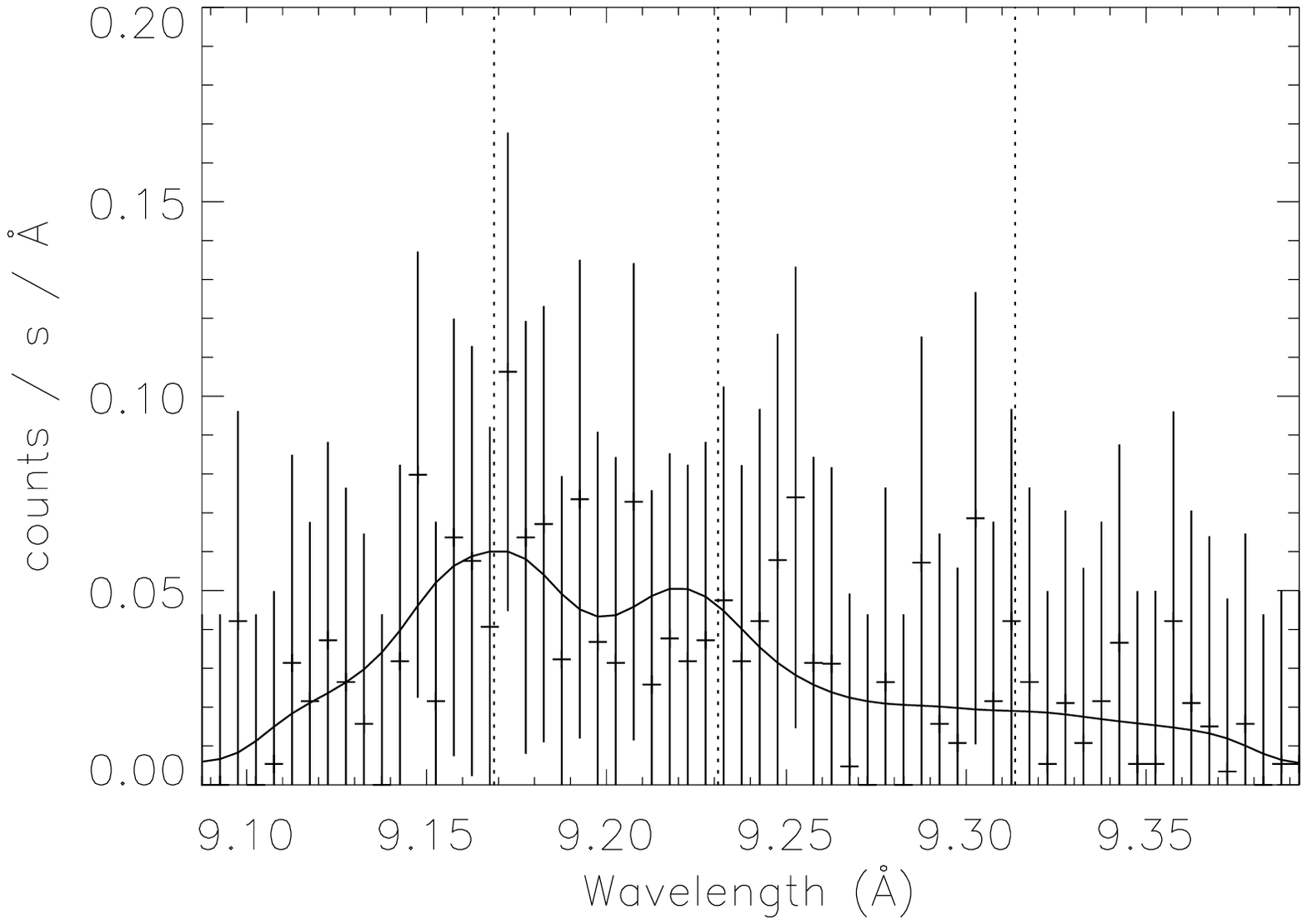}
  \end{center}
  \caption{MEG data and best-fit model for \ion{Mg}{11} in \io. The positive
    and negative first order data have been coadded.}
\label{IotaoriMgXI}
\end{figure}

\begin{figure}[p]
  \begin{center}
    \plotone{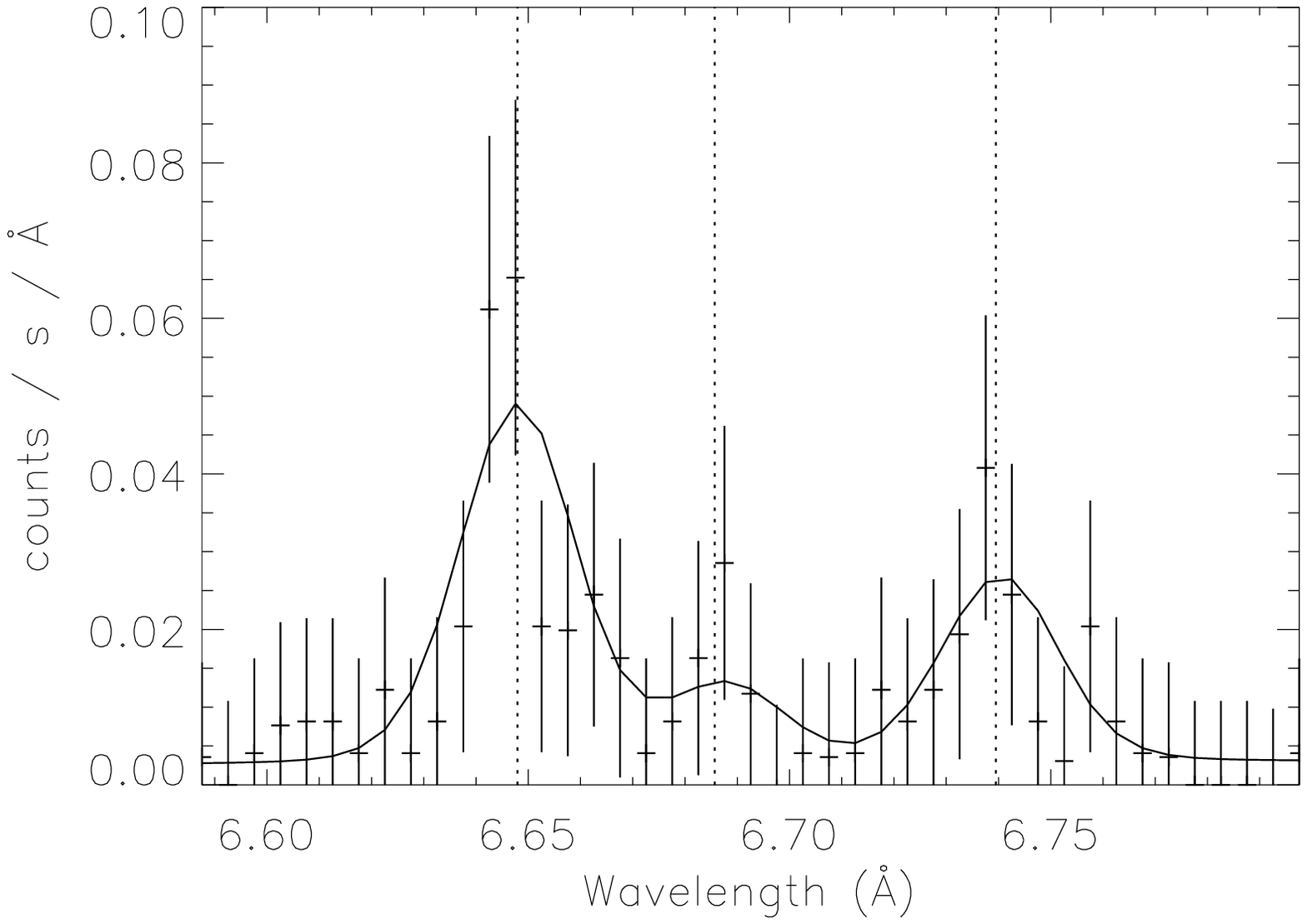}
  \end{center}
  \caption{MEG data and best-fit model for \ion{Si}{13} in \deltao. The
    positive and negative first order data have been coadded.}
\label{DeltaoriSiXIIIMEG}
\end{figure}

\begin{figure}[p]
  \begin{center}
    \plotone{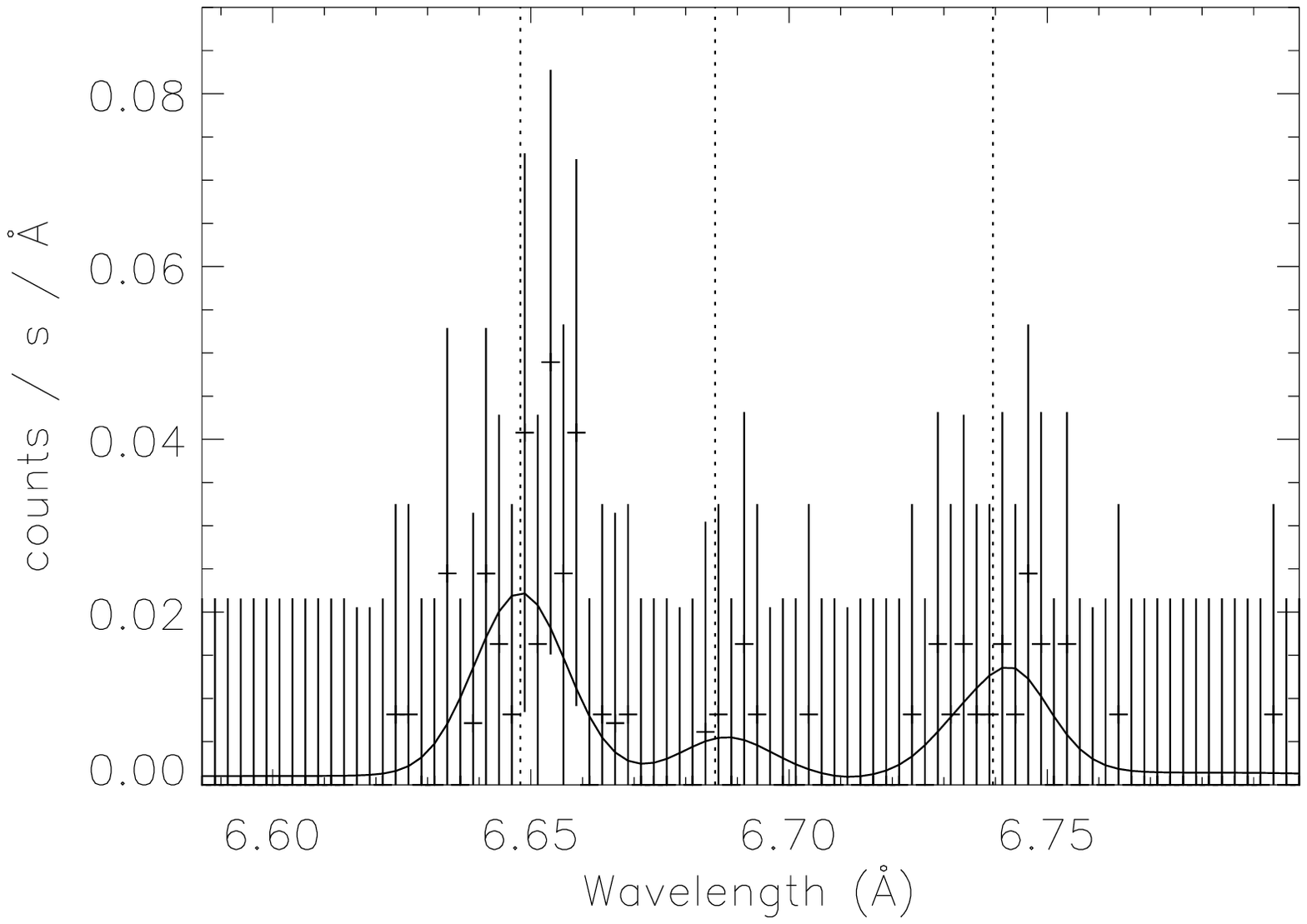}
  \end{center}
  \caption{HEG data and best-fit model for \ion{Si}{13} in \deltao. The
    positive and negative first order data have been coadded.}
\label{DeltaoriSiXIIIHEG}
\end{figure}

\begin{figure}[p]
  \begin{center}
    \plotone{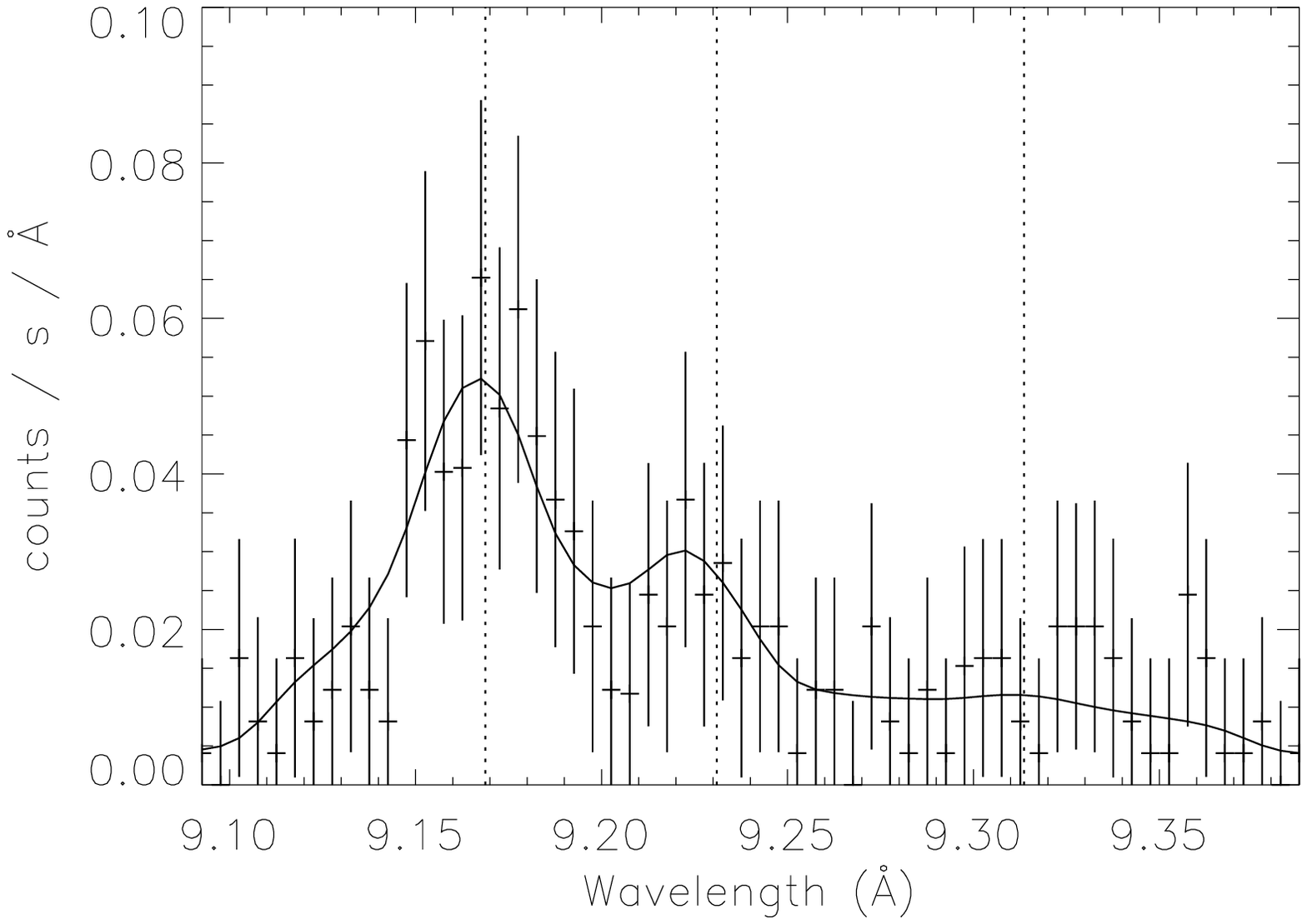}
  \end{center}
  \caption{MEG data and best-fit model for \ion{Mg}{11} in \deltao. The
    positive and negative first order data have been coadded.}
\label{DeltaoriMgXI}
\end{figure}

The results of the fits are summarized in Tables~\ref{TabHelikeProfile} and
\ref{TabHelikeGaussian}. The fits are plotted with the data in
Figures~\ref{ZetapupSXVMEG}-\ref{DeltaoriMgXI}. The
data have been rebinned for presentation purposes in some of the plots, but in
all cases the data were fit without rebinning. 

\begin{figure}[p]
  \begin{center}
    \epsscale{0.8}
    \plotone{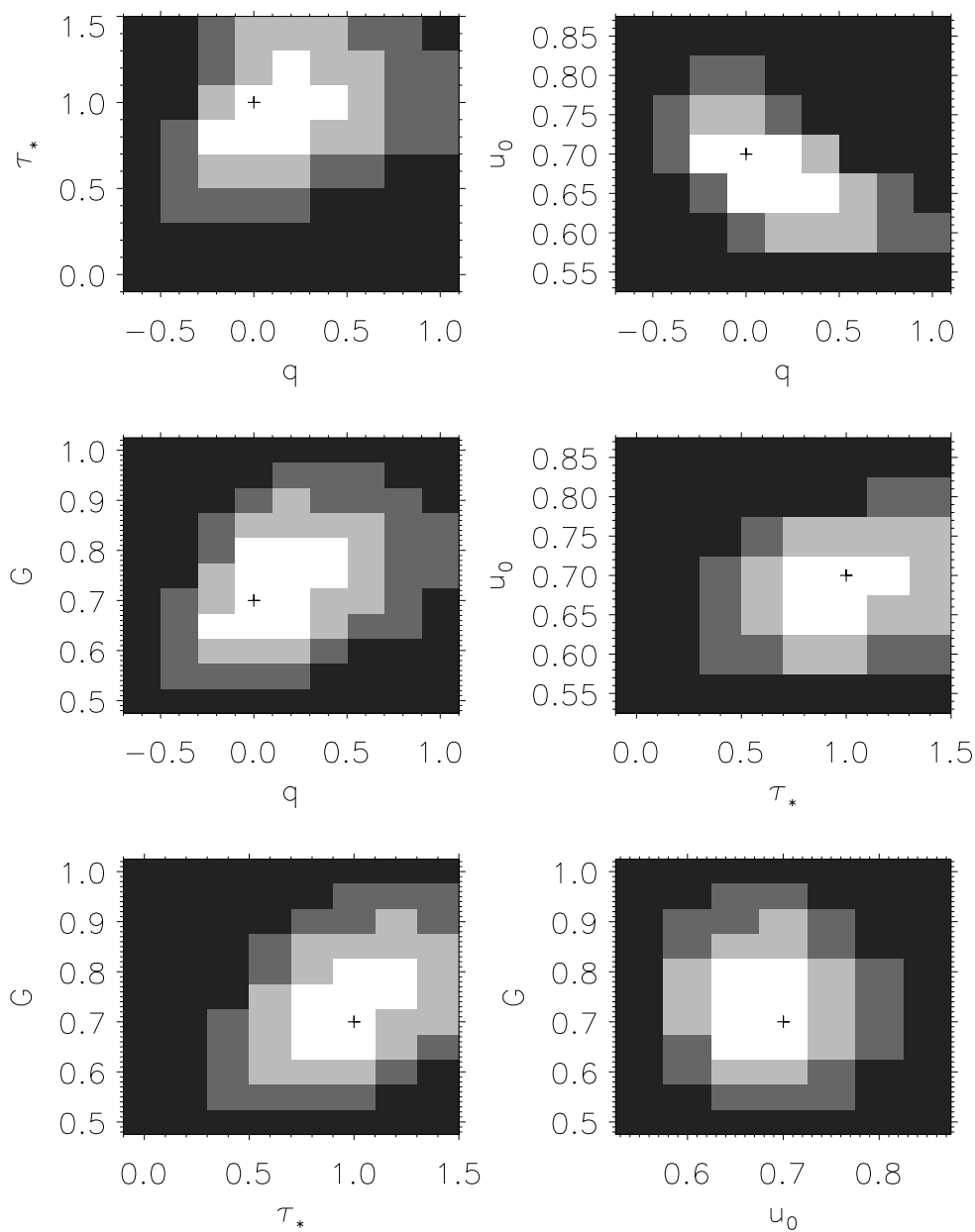}
  \end{center}
  \caption{Two dimensional plots of confidence intervals for fit parameters
    for \ion{Mg}{11} in \zp. The shades of grey represent 1, 2, and 3
    $\sigma$, or \(\Delta C<2.3,6.17,11.8\) (as appropriate for two degrees of
    freedom), and the cross represents the best
    fit. There is a moderate correlation of the fit parameters $q$ and
    \(u_0\), as one would expect. We have made similar plots for the other
    He-like profile fits (not shown) to look for correlations in fit
    parameters. These plots also show a moderate correlation between $q$ and
    \(u_0\).}
\label{ZetapupConfMgXI}
\end{figure}

We show two-parameter confidence interval plots for the profile fit to
\ion{Mg}{11} for \zp\, in Figure~\ref{ZetapupConfMgXI}. These confidence
intervals are qualitatively representative of our results for all the line
complexes; they demonstrate that there is a moderate correlation of the
parameters $q$ and \(u_0\) in the profile fits, and that the other parameters
are not strongly correlated. The correlation in $q$ and \(u_0\) is expected,
as both parameters influence the radial distribution of plasma, and therefore
both the $f/i$ ratio and the profile width.

The goodness of fit is tested by comparing the fit statistic to that obtained
from Monte Carlo simulations from the model. The percentage of 1000
realizations having $C$ less than the data is given in the tables of results.
These percentages can be thought of as rejection probabilities. 

The helium-like line profile fits generally are adequate to explain the
data; they are all formally statistically acceptable. The fact that the fits
can simultaneously account for the profile shape and the $f/i$ ratio indicates
that the values of \(u_0\) obtained are not an artifact of the profile
model. In other words, we can explain both the line ratios and profile shapes
with a single model for the radial distribution of X-ray emitting plasma.

The fit parameters obtained for the He-like profile fits are generally
consistent with those obtained in \citet{KCO03} and \citet{Cel06} from
non-helium-like line profile fits. The \({\cal R}\) ratios for the helium-like
line profile fits are also consistent with those measured in \citet{Kel01},
\citet{WC01}, \citet{CMWMC01}, and \citet{MCWMC02}. The values of \(u_0\) for
all four stars fall in the range \(0.6 < u_0 < 0.8\), or \(1.25 < R_0 <
1.67\). This is substantially closer to the star than the values of $u$
inferred in \citet{Kel01}, \citet{WC01}, and \citet{CMWMC01} from $f/i$
ratios. This reflects the difference between assuming a single radius of
formation as opposed to a distribution of radii.

We now consider the lower signal-to-noise complexes, which we fit with
Gaussians. For the \ion{Si}{13} lines in \io\, and \deltao, the \({\cal R}\)
ratio is not  strongly constrained, and in both cases the data are consistent
with \({\cal R} = {\cal R}_0\). The goodness of fit is formally acceptable in
both cases. If anything, it is surprising that the \({\cal R}\) ratio is not 
slightly lower in both cases, considering the values of \(u_0\)
measured for the \ion{Mg}{11} lines.

The \({\cal R}\) ratio measured in \ion{S}{15} in \zp\, is equivalent to a
value of \(R_0 = 1.1^{+0.4}_{-0.1}\), based on Figure~\ref{zetapuphelike}.
The \(1 \sigma\) upper limit to \(R_0\) is consistent with what is seen in
other lines and with the expectations of hydrodynamic models of wind shocks
\citep{FPP97, RO02}. The fit to these lines is formally acceptable.

The fit to the \ion{Si}{13} complex of \zo\, is poor. Because the positive and
negative first order MEG data look very different, we fit them separately in
addition to the joint fit. These additional fits are shown in
Figures~\ref{ZetaoriSiXIII+1} and \ref{ZetaoriSiXIII-1}. Part of the
difference in appearance is a result of the \ion{Si}{13} complex falling on a
chip gap in the negative first order, which reduces the effective area and
makes it uneven. However, even accounting for this there is a substantial
difference in the fit results for the two orders, both for \({\cal R}\) and
for $G$. It is possible to get a satisfactory fit using only the positive
first order MEG data, but fitting the negative first order by itself gives a
poor fit. Because the negative first order data for this complex falls on a
chip gap, cannot be fit well by a three Gaussian model, and has substantially
fewer counts than the positive first order, we consider it to be unreliable.

In Table~\ref{TabCompareMeasurements} we compare our fits for \ion{Si}{13} in
\zo\, and \ion{S}{15} in \zp\, to those of \citet{WC01} and \citet{CMWMC01},
respectively. There is not enough information in their original work to
directly compare their best fit model to ours; they do not give the velocity
broadenings or overall normalizations. We use their published
values of \({\cal R}\) and $G$ and find the best fit parameters for velocity
broadening and normalization. \citet{WC01} do not present their measurements
of $G$, but we infer from the temperature range they claim is allowed that
they measure $G$ in the range 0.8-2.0. We assume the best fit was in the
middle of this range, or $G=1.4$. In both cases, we also tried letting $G$ be
a free parameter, in order to test the validity of their \({\cal R}\)
measurements independently of any claims about $G$. For \ion{S}{15} in \zp, we
found that the best fit occurs with a substantially different value of $G$
than that reported by \citet{CMWMC01}.

Our measurement of the \({\cal R}\) ratio of \ion{Si}{13} in \zo\,
is significantly different than that of \citet{WC01}. Statistically, their
best-fit measurement has a $C$ value that is 11.7 greater than our best
fit. For one interesting parameter, this is excluded at more than \(3
\sigma\). The reported value of \({\cal R} = 1.2 \pm 0.5\) is also very
different than our measured value of \(2.8 \pm 0.8\). Two points should be
reiterated: first, when fitting all the data, we do not get a statistically
acceptable fit, but the positive first order MEG data can be well-fit, and
this fit has an \({\cal R}\) ratio which is comparable to the value we measure
using all the data; furthermore, whether we use all the data or exclude the
questionable negative first order MEG data from the fit, we get essentially
the same result. Second, we are using essentially the same model as
\citet{WC01}, but merely measure very different parameter values, even when
fitting exactly the same data. This may stem from the fact that \citet{WC01}
used very early versions of the CIAO tools (J. P. Cassinelli, private
communication).

Our measurement of the \({\cal R}\) ratio of \ion{S}{15} in \zp\, is somewhat
different than that of \citet{CMWMC01}. We also found that the if we fix
\({\cal R}\) to the value they reported, the best fit value of $G$ is
substantially different than their measurement. Although their best fit model
with \({\cal R} = 0.61\) is not excluded at the \(1 \sigma\) level, the model
based on their measured value of $G$ has a value of $C$ which is greater than
that of our best fit by 9.5, despite the fact that we both fit a three
Gaussian model. Although we do not exclude their best-fit value of \({\cal
  R}\) at \(1 \sigma\), it is also puzzling that our range of fit values
should be significantly different from that of the previous work.

\section{Discussion \label{sec:discussion}}

We have used the $f/i$ ratios of He-like triplets in conjunction with their
line profiles to constrain the radial distribution of X-ray emitting plasma in
O stars. Our results are consistent with the results of \citet{KCO03} and
\citet{Cel06} in the sense that the spatial distribution we infer from $f/i$
ratios (additionally constrained in some cases by line profile fitting) is
consistent with these authors' results from fitting line profiles to high
signal-to-noise individual lines. 

\begin{figure}[p]
  \begin{center}
    \plotone{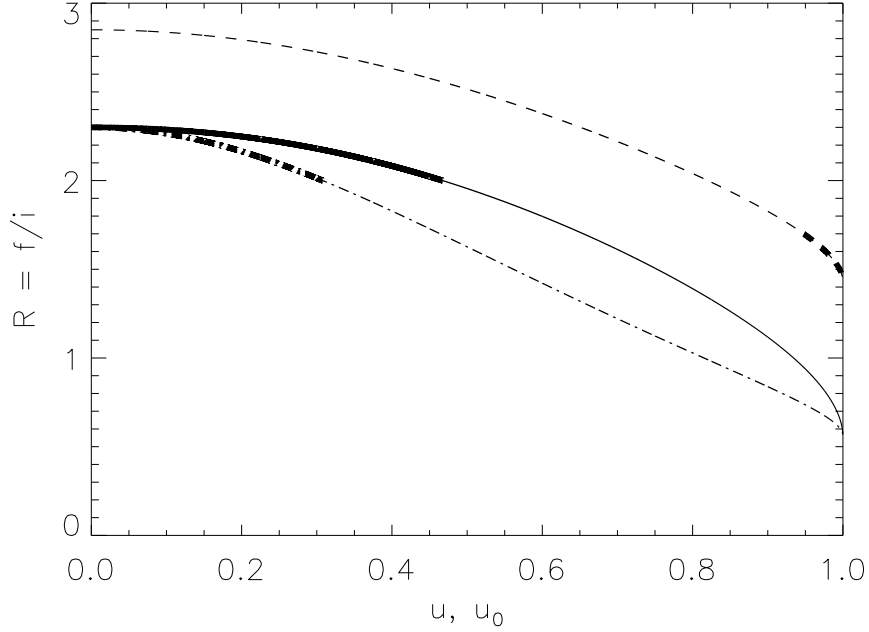}
  \end{center}
  \caption{Comparison of measurements and calculations for \ion{Si}{13} in
    \zo. Calculations are thin lines, and measurements are thickened over the
    allowed range of \({\cal R}\). The solid line shows \(\overline{\cal
      R}(u_0)\) from this work; the dash-dot line shows the \({\cal R}(u)\)
    for a single radius, but using an averaged value of the TLUSTY UV flux; and
    the dashed line shows the calculations and measurements of
    \citet{WC01}. Note that the range shown by the thickened lines represents
    the allowed range of measured \({\cal R}\) or \(\overline{\cal R}\)
    values, and does not represent the physical extent of the X-ray emitting
    plasma. In the case of \({\cal R}\) the model assumes a single radius of
    formation, while for \(\overline{\cal R}\) the value of \(u_0\) inferred
    corresponds to the minimum radius for X-ray emission. The fact that the
    allowed range of \({\cal R}\) graphically mimics the distribution of
    plasma radii for the upper limit value to \(u_0\) is a coincidence.}
  \label{zetaoricomparison}
\end{figure}

Our results for \ion{Si}{13} in \zo\, are different from the initial analysis
which claimed that the location of the emitting plasma was extremely close to
the star.  These differences are due both to our assessment of the relative
line fluxes and to our modeling of the line formation.
Table~\ref{TabCompareMeasurements} shows a comparison of our measurements and
inferred radii of formation; we find that the Si XIII is at least 1.1 stellar
radii above the photosphere (\(R_0 / R_* = 2.1\)). Part of the difference in
inferred radii originates in our different calculations of the radial
dependence of \({\cal R}\). This is illustrated in
Figure~\ref{zetaoricomparison}, where we plot our calculations and
measurements of \(\overline{\cal R}(u_0)\) and compare them to the
calculations and measurements of \({\cal R}(u)\) from \citet{WC01}. (It is
important to note that the range of radii indicated on the plot by the
thickened lines refers to that allowed by the statistical error in the
measurement of \({\cal R}\), and not to a physical extent of the X-ray
emitting plasma). We also show what our inferred radius of formation would be
if we inferred a single radius from our measured value of \({\cal R}\) instead
of an onset radius \(R_0\) in a distributed model (assuming that an averaged
value of the photospheric UV flux could be used). This is intended to make it
clear that the major sources of disagreement are the actual \({\cal R}\)
measurements and the UV fluxes of the adopted model atmospheres. In fact, even
taking the reported upper limit on \cal{R} (of 1.7) from WC2001, and assuming
a single radius of formation (dash-dot curve in
Figure~\ref{zetaoricomparison}), our analysis shows that the formation radius
is consistent with values larger than \(2 R_*\). Although we also assume a
spatial distribution of X-ray emitting plasma, this does not contribute to the
new, larger formation radii. 

\begin{figure}[p]
  \begin{center}
    \plotone{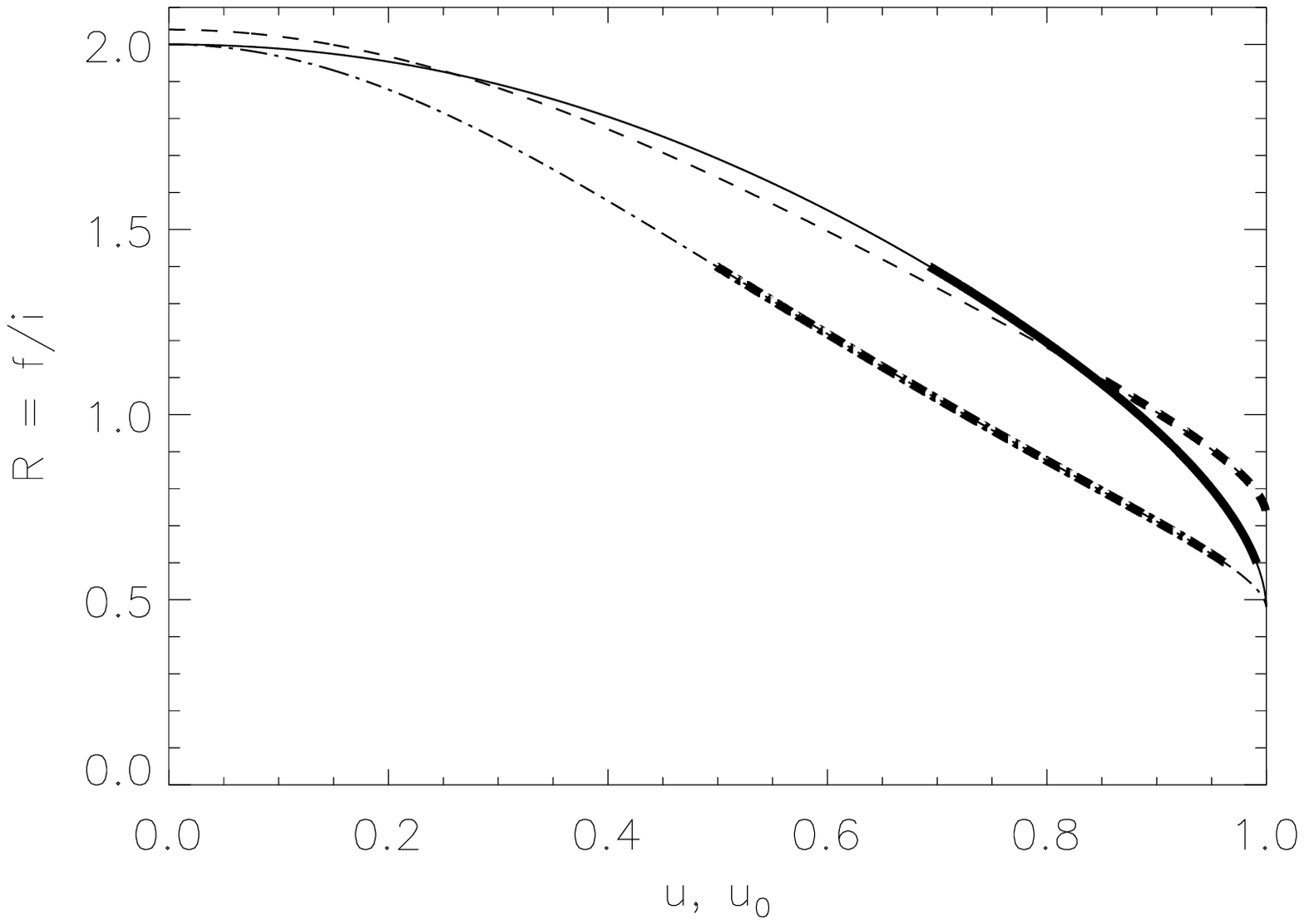}
  \end{center}
  \caption{Same as Figure~\ref{zetaoricomparison}, but for \ion{S}{15} for
    \zp, and we are comparing our work to \citet{CMWMC01}.}
  \label{zetapupcomparison}
\end{figure}

We make a similar comparison with the earlier results \citep{CMWMC01} for
\ion{S}{15} for \zp\, in Table~\ref{TabCompareMeasurements} and
Figure~\ref{zetapupcomparison}. In this case the measured range of allowed
values of \({\cal R}\) is different but overlapping. The different measured
range of \({\cal R}\) combined with a somewhat higher model photospheric UV
flux leads us to infer a minimum radius of formation as large as \(1.5
R_*\); however the allowed range of minimum radii extends down to nearly the
photosphere, in agreement with the results of \citet{CMWMC01}. The upper range
of allowed minimum radii is reasonable in the context of stellar wind models
for X-ray-emitting plasma formation, but the lower range is certainly
not. While the difference between our measurements and calculations and those
of \citet{CMWMC01} is not great, it is enough to allow that the
\ion{S}{15} emission could reasonably be produced in a wind shock model.

These results obviate the need for any kind of two-component model for the
origin of X-ray emission in O stars, as suggested by \citet{WC01},
\citet{CMWMC01}, and \citet{MW06}. For the \ion{Si}{13} line in \zo\, the
range of acceptable minimum radii of formation we infer are quite reasonable
in the wind-shock paradigm. For the \ion{S}{15} line in \zp, the upper end of
the range of acceptable minimum radii of formation we infer is acceptable in the
wind-shock paradigm, although the lower end of the range is not. Taken
together, we can say that the wind-shock paradigm is consistent with these
data; we do not exclude the possibility that \ion{S}{15} in \zp\, is formed
very close to the star or is formed in a process outside of the wind-shock
paradigm, but we do not require this. We note that numerical simulations of
the line-driven instability show that large shock velocities, and therefore
hot plasma, occur quite deep in the wind, almost as soon as the damping
effects of the diffuse radiation field are overcome by the onset of the
instability growth \citep{RO02}. While hybrid wind-coronal mechanisms are not
excluded by the data, there is nothing in the X-ray spectral data that
requires such complex models, and the principle of Occam's razor leads us to
suggest that it is more reasonable to assume a wind shock origin for all
X-rays from the O stars we are studying, if it is possible to explain the data
this way. Another argument against inferring extremely small formation radii
for these two ions is the lack of any evidence, either from line profiles or
from $f/i$ ratios, of the presence of emission from lower ion stages at these
very small radii, as would be expected from the rapid radiative cooling of
plasma containing \ion{S}{15} or \ion{Si}{13} at the densities expected this
far down in the wind.  

It should be noted that there is one other claim in the
literature of an anomalously low $f/i$ ratio measurement requiring
an X-ray production mechanism outside of the standard wind shock
paradigm. \citet{Wel04} find evidence for this in their analysis of the X-ray
spectrum of Cyg OB2 8A; in this case the basis for their claim is emission
from \ion{S}{15} and \ion{Ar}{17}. However, these ions' \nspec{2}{3}{S}{1}
\(\rightarrow\) \nspec{2}{3}{P}{J} transitions are in the Lyman continuum,
where results are very sensitive to model atmosphere uncertainties, so
the inferred radii are subject to substantial uncertainties. Furthermore, the
data have very low signal-to-noise.

The characteristic optical depths we measure from profile fits are
substantially smaller than one would expect, given the published mass-loss
rates. Detailed calculations of the expected values of \(\tau_*\) are beyond
the scope of this paper, but it is safe to say that we would expect to see
characteristic optical depths at least of order a few at 9 \AA; our
measurements for \ion{Mg}{11} give \(\tau_* = 1\) for \zp, and less for the
other stars. However, \ion{Si}{13} and \ion{Mg}{11} give poorer constraints on
the optical-depth/mass-loss-rate discrepancy than longer wavelength lines,
such as \ion{O}{8} and \ion{N}{7} Lyman \(\alpha\), where the photoelectric
absorption cross-section per unit mass is higher, so that \(\tau_*\) is larger
and produces a more asymmetric profile.

\citet{WC01}, \citet{CMWMC01}, and \citet{MCWMC02} compare the radii they
infer from measurements of $f/i$ ratios in He-like triplets to the radii of
optical depth unity, \(R_1\), for the wavelength at which that He-like ion
emits. These values of \(R_1\) were calculated using mass-loss rates from the
literature and assumed a smooth wind density. They claim that the inferred
radii correspond roughly to \(R_1\), so that we are observing plasma at the
closest point to the star where we can see it. \citet{Kel01} make a similar
conjecture. Table~\ref{TabROne} compares the values of \(R_1\) from these
papers to those derived using the methodology of this paper. Several lines
show evidence for emission from inside the predicted \(R_1\). This is in
agreement with the low values of \(\tau_*\) we have measured, as well as the
measurements of \citet{KCO03} and \citet{Cel06}. There is now mounting
evidence from analysis of unsaturated UV line profiles that the literature
mass-loss rates of O stars may be too high by at least a factor of a few
\citep{MFSH03, Hel03, BLH05, FMP06}. In addition, porosity may reduce the
effective X-ray optical depths of O star winds \citep{FOH03, OFH04, OFH06,
  OC06}.

\section{Conclusions \label{sec:conclusions}}
We have investigated the effect of a radially distributed plasma on the
forbidden-to-intercombination line ratio in helium-like triplets, as well as
variations in the exciting photospheric flux as a function of Doppler shift
throughout the wind. We find that the fact that the plasma is likely
distributed over a range of radii and Doppler shifts allows us to use an
averaged value of the photospheric continuum instead of accounting for it in
detail. We also find that the value of \(R_0\) derived assuming
a distribution of radii is substantially smaller than the value of $R$ derived
assuming a single radius.

We have used the $f/i$ ratio of helium-like triplets to constrain the radial
distribution of X-ray emitting plasma in four O-type stars. We find that the
minimum radius of emission is typically \(0.6 < u_0 < 0.8\), or
\(1.25 < R_0 / R_* < 1.67\) with the emission extending beyond this initial
radius with either a constant filling factor or one that increases slightly
with radius. This is consistent with the results of line profile fits using
the model of \citet{OC01} \citep{KCO03, Cel06}. However, some of the minimum
radii of formation are well inside the radius of optical depth unity
calculated using the mass-loss rates in the literature, implying that either
the effective opacities are lower \citep[e.g. due to porosity
  effects][]{FOH03, OFH04, OFH06, OC06} or the mass-loss rates are lower than
the literature values \citep{MFSH03, Hel03, BLH05, FMP06} or both. We also
measure low values of the characteristic optical depth \(\tau_*\) compared to
what one would expect based on the literature mass-loss rates, which is
consistent with the same conclusions.

We find that there is no evidence for anomalously low $f/i$ ratios in high-$Z$
species. Our measurements do not require X-ray emission orignating from too
close to the star to have sufficiently strong shocks, nor do we need to posit
the existence of a magnetically confined corona. This conclusion is based
partly on different measured values of $f/i$ ratios and partly on higher
photospheric UV fluxes on the blue side of the Lyman edge in the more recent
TLUSTY model spectra. 

We have fit He-like emission line complexes with profile models that
simultaneously account for profile shapes and line ratios. These models
constrain the radial distribution of plasma both through the line ratio and
the profile parameters \(u_0\) and $q$. We find that they are
capable of producing good fits to the data, showing that the information
contained in the line ratios and profile shapes are mutually consistent.

\acknowledgements

We thank J. Cassinelli and the anonymous referee for many useful comments that
led to improvements in the manuscript. MAL acknowledges NASA grant NNG04GL76G.
DHC acknowledges NASA contract AR5-6003X to Swarthmore College through the
Chandra X-ray Center.


\bibliography{xray-ostar}

\clearpage
\begin{deluxetable}{cccccc}
\tablecaption{Parameters adopted for He-like triplets \label{TabHeParams}}
\tablehead{
\colhead{Ion} & 
\colhead{$f$ (\nspec{2}{3}{S}{1} \(\rightarrow\) \nspec{2}{3}{P}{J})
  \tablenotemark{a}} &  
\colhead{\(\phi_c\) (\({\mathrm s^{-1}}\)) \tablenotemark{b}} & 
\colhead{\({\cal R}_0\)\tablenotemark{c}} & 
\colhead{\(G(T_{max})\)\tablenotemark{c}}
}
\startdata
\ion{S}{15}  & 0.0507 & 9.16e5 & 2.0 & \nodata \\
\ion{Si}{13} & 0.0562 & 2.39e5 & 2.3 & 0.68 \\
\ion{Mg}{11} & 0.0647 & 4.86e4 & 2.7 & 0.71 \\
\ion{Ne}{9}  & 0.0700 & 7.73e3 & 3.1 & 0.74 \\
\ion{O}{7}   & 0.0975 & 7.32e2 & 3.7 & 0.90 \\
\ion{N}{6}   & 0.1136 & 1.83e2 & 5.3 & 0.88 \\
\enddata
\tablenotetext{a}{Oscillator strengths are from CHIANTI \citep{Del97, Yel03}.}
\tablenotetext{b}{\(\phi_c\) are from \citet*{BDT72}.}
\tablenotetext{c}{ \({\cal R}_0\) and \(G\, (T_{max})\) are from
  \citet{PMDRK01}, except \ion{S}{15}, which is from \citet*{BDT72}.}
\end{deluxetable}

\begin{deluxetable}{ccccc}
\tablecaption{Adopted stellar parameters \label{TabStarParams}}
\tablehead{
\colhead{Star} & \colhead{Spectral type \tablenotemark{a}} & 
\colhead{\(T_{eff}\) (kK) \tablenotemark{b}} & 
\colhead{\(\log g\) (\(\mathrm {cm\, s^{-2}}\)) \tablenotemark{b}} & 
\colhead{\(v_{\infty}\) (\({\mathrm {km\,s^{-1}}}\))\tablenotemark{c}}
}
\startdata
\zp & O4\, I & 42.5 & 3.75 & 2485 \\
\zo & O9.5\, I & 30.0 & 3.25 & 1860 \\
\io & O9\, III & 35.0 & 3.50 & 2195 \\
\deltao & O9.5\, II & 32.5 & 3.25 & 1995 \\
\enddata
\tablenotetext{a}{Spectral types are given for reference and are taken from
  the Garmany values reported in Table~1 of \citet{LL93}.}
\tablenotetext{b}{Effective temperatures and surface gravities are the values
  on the TLUSTY O star grid that are the closest approximations to the values
  used in \citet{LL93}.}
\tablenotetext{c}{Terminal velocities are taken from \citet*{PBH90}.}
\end{deluxetable}

\begin{deluxetable}{cccccccc}
\tablecaption{Comparison of He-like ratio calculations 
  \label{TabCompareTLUSTYKurucz}}
\tablehead{
\colhead{} & \colhead{} & \colhead{} & 
\colhead{\(\overline{\frac{H_{\nu}}{E}}\)\tablenotemark{b}} & 
\colhead{$P$ \tablenotemark{c}} &
\colhead{\({\cal R}_{ph}/{\cal R}_0\)} & 
\colhead{\({\cal R}_0\) \tablenotemark{d}} &
\colhead{\({\cal R}_{ph}\) \tablenotemark{e}}
}
\startdata
\zo & \ion{Si}{13} & this work & 1.97  & 3.09  & 0.244 & 2.3  & 0.56 \\
    & & WC01 \tablenotemark{a} & 0.633 & 0.993 & 0.502 & 2.85 & 1.43 \\
\hline
\zp & \ion{S}{15}  & this work & 8.71 & 3.21 & 0.238 & 2.0  & 0.48 \\ 
    &  & C01 \tablenotemark{a} & 4.95 & 1.82 & 0.355 & 2.04 & 0.72 \\
\enddata
\tablenotetext{a}{We used Kurucz model atmospheres to reproduce these authors'
  calculations. We assumed that \zp\, was represented by a model with 
  \(T_{eff} = 40 \mathrm{kK}\) and \(\log g = 4.0\) and \zo\, by a model with
  \(T_{eff} = 30 \mathrm{kK}\) and \(\log g = 3.5\).}
\tablenotetext{b}{The photospheric UV flux, \(\overline{\frac{H_{\nu}}{E}}\),
  is given in units of \(10^{7}\, \mathrm{photons\: cm^{-2}\: s^{-1}\:
    Hz^{-1}}\).} 
\tablenotetext{c}{\(P \equiv \phi_* / \phi_c\) is discussed in
  Equations~\ref{eqn:rofr} and \ref{eqn:phistar}.}
\tablenotetext{d}{For WC01 and C01 we used the \({\cal R}_0\) values shown on
  their plots.}
\tablenotetext{e}{ Our calculations for  \({\cal R}_{ph}\) (which is the value
  of \({\cal R}\) near the photosphere) using the Kurucz model atmospheres
  agree with the figures of WC01 and C01.}
\tablecomments{In this table we compare the adopted photospheric UV flux and
  the He-like triplet ratio calculations of \citet{WC01} and \citet{CMWMC01}
  to those in this work.}
\end{deluxetable}

\clearpage

\begin{deluxetable}{cccccccccccc}
\tablewidth{0pt}
\tabletypesize{\footnotesize}
\tablecaption{Parameters for He-like profile fits \label{TabHelikeProfile}} 
\tablehead{
\colhead{Star} & \colhead{Ion} & \colhead{$q$} & \colhead{\(\tau_*\)} & 
\colhead{\(u_0\)} & \colhead{\(R_0\) \tablenotemark{a}} & \colhead{$G$} & 
\colhead{\({\cal R}\) \tablenotemark{b}} & \colhead{Flux \tablenotemark{c}} & 
\colhead{$C$} & \colhead{Bins} & \colhead{MC \tablenotemark{d}}
}
\startdata
$\zeta$ Pup  & \ion{Mg}{11} & $0.0^{+0.4}_{-0.2}$ & $1.0^{+0.4}_{-0.4}$ &
$0.70^{+0.05}_{-0.05}$ & $1.43$ & $0.70^{+0.15}_{-0.10}$ & $0.41$ &
$17.7^{+0.9}_{-0.9}$ & 135.3 & 136 & 39.1 \\
             & \ion{Si}{13} & $0.0^{+0.6}_{-0.4}$ & $0.6^{+0.4}_{-0.2}$ &
$0.70^{+0.05}_{-0.10}$ & $1.43$ & $1.05^{+0.15}_{-0.15}$ & $0.90$ &
$11.9^{+0.7}_{-0.7}$ & 116.2 & 98 & 84.7 \\
$\zeta$ Ori  & \ion{Mg}{11} & $-0.4^{+1.0}_{-0.2}$ & $0.2^{+0.2}_{-0.2}$ &
$0.6^{+0.10}_{-0.1*}$ & $1.67$ & $1.05^{+0.05*}_{-0.2}$ & $0.82$ & 
$6.5^{+0.5}_{-0.6}$ & 267.2 & 240 & 73.4 \\
$\iota$ Ori  & \ion{Mg}{11} & $-0.8^{+0.2}_{-0.0}$ & $0.0^{+0.2}_{-0.0}$ &
$0.75^{+0.05}_{-0.10}$ & $1.33$ & $0.90^{+0.20*}_{-0.25}$ & $0.72$ & 
$3.5^{+0.5}_{-0.5}$ & 266.6 & 256 & 80.2 \\
$\delta$ Ori & \ion{Mg}{11} & $-0.8^{+0.4}_{-0.0}$ & $0.0^{+0.2}_{-0.0}$ &
$0.80^{+0.05}_{-0.10}$ & $1.25$ & $0.60^{+0.25}_{-0.10*}$ & $0.75$ & 
$4.0^{+0.5}_{-0.5}$ & 123.9 & 124 & 18.2 \\
\enddata
\tablenotetext{a}{\(R_0\) is given in units of the stellar radius; it is
  calculated from \(R_0 = 1/ u_0\), and retains an extra digit to avoid
  rounding error.} 
\tablenotetext{b}{\({\cal R}\) is reported without an uncertainty because it
  is the value of the $f/i$ ratio calculated by the best fit model.}
\tablenotetext{c}{Flux is given in units of \(10^{-5}\,\mathrm{photons\:
    cm^{-2}\: s^{-1}}\).} 
\tablenotetext{d}{MC is the percentage of Monte Carlo realizations of the
  model having C less than the data does for that model.}
\tablecomments{Errors are $2\sigma$, or $\Delta C = 4$ for one degree of
  freedom. Asterisks indicate parameters that were still within $2\sigma$
  at the edge of the fit range.}
\end{deluxetable}

\clearpage

\begin{deluxetable}{cccccccccc}
\tablewidth{0pt}
\tabletypesize{\footnotesize}
\tablecaption{Parameters for He-like Gaussian fits \label{TabHelikeGaussian}} 
\tablehead{
\colhead{Star} & \colhead{Ion} & \colhead{\(\sigma_v / c\, (10^{-3})\)} &
\colhead{\({\cal R}\)} & \colhead{$G$} & \colhead{Flux \tablenotemark{a}} & 
\colhead{$C$} & \colhead{Bins} &  \colhead{MC \tablenotemark{b}} &
\colhead{\(R_0\)}
}
\startdata
\zp & \ion{S}{15} & $2.4^{+0.4}_{-0.4}$ & $1.0^{+0.4}_{-0.4}$ &
$0.9^{+0.2}_{-0.2}$ & $3.1^{+0.3}_{-0.3}$ & 191.2 & 216 & 78.4 & 
$1.1^{+0.4}_{-0.1}$ \\
\zo & \ion{Si}{13} \tablenotemark{c} & $2.4^{+0.2}_{-0.0}$ & $2.8\pm 0.8$ &
$1.2^{+0.2}_{-0.1}$ & $2.45\pm 0.15$ & 432.1 & 496 & 97.9 & $\geq 2.1$ \\
& \tablenotemark{d} & $2.0\pm 0.2$ & $\geq 2.8$ &
$0.9\pm 0.2$ & $2.4\pm 0.4$ & 57.6 & 59 & 35.3 & $\infty$ \\
& \tablenotemark{e} & $3.0^{+0.8}_{-0.6}$ & $\geq 1.6$ &
$2.0^{+1.0*}_{-0.6}$ & $2.4^{+0.6}_{-0.5}$ & 85.6 & 59 & 98.8 & $\geq 1.4$ \\
\io & \ion{Si}{13} & $2.8\pm 0.4$ & $2.8^{+0.6}_{-0.8}$ &
$1.6^{+0.4}_{-0.2}$ & $1.54\pm 0.24$ & 305.1 & 532 & 84.9 & $\geq 3.2$ \\
\deltao & \ion{Si}{13} & $1.2^{+0.0}_{-0.2}$ & $2.2^{+1.0}_{-0.4}$ &
$0.7\pm 0.1$ & $1.88\pm 0.16$ & 214.0 & 258 & 62.9 & $\geq 2.2$ \\
\enddata
\tablenotetext{a}{Flux is in units of \(10^{-5}\,\mathrm{photons\: cm^{-2}\:
    s^{-1}}\).} 
\tablenotetext{b}{MC is the percentage of Monte Carlo realizations of the
  model having C less than the data does for that model.}
\tablenotetext{c}{Combined fit to positive and negative first order HEG and
  MEG data.}
\tablenotetext{d}{Fit to positive first order MEG data only.}
\tablenotetext{e}{Fit to negative first order MEG data only.}
\tablecomments{Errors are $1\sigma$, or $\Delta C = 1$. Asterisks indicate
  parameters that were still within $1\sigma$ at the edge of the fit range.}
\end{deluxetable}

\clearpage

\begin{deluxetable}{ccccccccc}
\tablecaption{Comparison of fit parameters \label{TabCompareMeasurements}}
\tablehead{
\colhead{} & \colhead{Star} & \colhead{Ion} & 
\colhead{\(\sigma_v / c\, (10^{-3})\)} & \colhead{\({\cal R}\)} & 
\colhead{$G$} & \colhead{$C$} & \colhead{bins} & 
\colhead{\(R_0\), $R$ \tablenotemark{d}}
}
\startdata
This work & \zp & \ion{S}{15}  & 2.4 & 1.0  & 0.9  & 191.2 & 216 &
\(1.1^{+0.4}_{-0.1}\)\\
C01a\tablenotemark{a} & &      & 2.8 & 0.61 & 2.06 & 200.7 &     & 
\(< 1.2\)\\ 
C01b\tablenotemark{b} & &      & 2.4 & 0.61 & 0.9  & 192.0 &     & \\
\hline
This work & \zo & \ion{Si}{13} & 2.4 & 2.8  & 1.2  & 432.1 & 496 & 
\(\geq 2.1\) \\
WC01 \tablenotemark{c} & &     & 2.2 & 1.2  & 1.4  & 443.8 &     & 
\(< 1.08 \) \\
\enddata
\tablenotetext{a}{For C01a we used the published value of \({\cal R}\) and
  $G$.}
\tablenotetext{b}{For C01b we used the published value of \({\cal R}\) and the
  best fit value for $G$.}
\tablenotetext{c}{For WC01 the published value of $G$ was the best fit,
  assuming their value of \({\cal R}\).}
\tablenotetext{d}{For our work this column gives the inferred minimum radius
  of formation \(R_0\), while for the previous work this column gives the
  inferred radius of formation $R$.}
\tablecomments{In this table we compare our fit parameters to those of
  \citet{CMWMC01} and \citet{WC01} for two line complexes. In all cases, we
  used the best fit \(\sigma_v\) and normalization.}
\end{deluxetable}

\clearpage

\begin{deluxetable}{cccc}
\tablecaption{Comparison of \(R_1\) to \(R_0\) \label{TabROne}}
\tablehead{
  \colhead{Star} & \colhead{Ion} & \colhead{\(R_1\)} & 
\colhead{\(R_0\) \tablenotemark{a}}
}
\startdata
\zp     & \ion{Mg}{11} & 2.5 & \(1.43\pm 0.10\) \\
        & \ion{Ne}{9}  & 5   & \(< 2.5\)  \\
        & \ion{O}{7}   & 4   & \(< 4\) \\
\zo     & \ion{Mg}{11} & 1.5 & \(1.67^{+0.33}_{-0.24}\)\\
        & \ion{Ne}{9}  & 2.8 & \(< 2.22\)  \\
        & \ion{O}{7}   & 2.2 & \(< 2.85\) \\
\deltao & \ion{Mg}{11} & 1.2 & \(1.25^{+0.18}_{-0.07}\)\\
        & \ion{Ne}{9}  & 2.1 & \(< 2.22\)  \\
        & \ion{O}{7}   & 2.8 & \(< 3.33\) \\
\enddata
\tablenotetext{a}{\(R_0\) is measured using an He-like line profile fit for
  \ion{Mg}{11} (see Table~\ref{TabHelikeProfile}). Upper limits for \ion{O}{7}
  and \ion{Ne}{9} are derived from upper limits to the \(f / i\) ratio of
  \({\cal R} < 0.1\) and \({\cal R} < 0.2\), respectively, which are taken to
  be representative for all three stars.}
\tablecomments{In this table we compare the radius of optical depth unity
  \(R_1\) calculated by \citet{WC01}, \citet{CMWMC01}, and \citet{MCWMC02} to
  measurements of \(R_0\), the minimum onset radius for X-ray emission.}
\end{deluxetable}

\clearpage


\clearpage

\clearpage

\clearpage


\end{document}